\documentclass[11pt]{aastex631}

\hypersetup{linkcolor=red,filecolor=cyan,urlcolor=magenta}

\usepackage{amsmath}
\usepackage{amssymb}
\usepackage{wasysym}
\usepackage{multirow}
\usepackage{longtable}
\usepackage{natbib}
\usepackage{colortbl}
\usepackage{enumitem}

\definecolor{updatecolor}{rgb}{0.5,0.0,0.8}

\useunder{\uline}{\ul}{}

\newcommand\latex{La\TeX}

\newcommand{\tempo}{\texttt{tempo}}
\newcommand{\tempotwo}{\texttt{tempo2}}
\newcommand{\pint}{\texttt{PINT}}
\newcommand{\enterprise}{\texttt{ENTERPRISE}}
\newcommand{\temponest}{\texttt{TEMPONEST}}
\newcommand{\ptmcmc}{\texttt{PTMCMCSampler}}

\newcommand{\updated}[1]{{#1}}

\begin{document}

\title{\pint{}: Maximum-likelihood estimation of pulsar timing  noise parameters}

\author[0000-0002-2820-0931]{Abhimanyu Susobhanan}
\affiliation{Max-Planck-Institut f{\"u}r Gravitationsphysik (Albert-Einstein-Institut), Callinstra{\ss}e 38, D-30167, Hannover, Germany\\
Leibniz Universit{\"a}t Hannover, D-30167, Hannover, Germany}
\affiliation{Center for Gravitation, Cosmology and Astrophysics, Department of Physics, University of Wisconsin-Milwaukee,\\ P.O. Box 413, Milwaukee, WI 53201, USA}

\author[0000-0001-6295-2881]{David L. Kaplan}
\affiliation{Center for Gravitation, Cosmology and Astrophysics, Department of Physics, University of Wisconsin-Milwaukee,\\ P.O. Box 413, Milwaukee, WI 53201, USA}

\author[0000-0003-0638-3340]{Anne M. Archibald}
\affiliation{Newcastle University, NE1 7RU, UK}

\author[0000-0001-5373-5914]{Jing Luo}
\altaffiliation{Deceased}
\affiliation{Department of Astronomy \& Astrophysics, University of Toronto, 50 Saint George Street, Toronto, ON M5S 3H4, Canada}

\author[0000-0002-5297-5278]{Paul S. Ray}
\affiliation{Space Science Division, Naval Research Laboratory, Washington, DC 20375-5352, USA}

\author[0000-0001-5465-2889]{Timothy T. Pennucci}
\affiliation{Institute of Physics and Astronomy, E\"{o}tv\"{o}s Lor\'{a}nd University, P\'{a}zm\'{a}ny P. s. 1/A, 1117 Budapest, Hungary}

\author[0000-0001-5799-9714]{Scott M. Ransom}
\affiliation{National Radio Astronomy Observatory, 520 Edgemont Road, Charlottesville, VA 22903, USA}

\author[0000-0001-5134-3925]{Gabriella Agazie}
\affiliation{Center for Gravitation, Cosmology and Astrophysics, Department of Physics, University of Wisconsin-Milwaukee,\\ P.O. Box 413, Milwaukee, WI 53201, USA}

\author[0000-0001-5645-5336]{William Fiore}
\affiliation{Department of Physics and Astronomy, West Virginia University, P.O. Box 6315, Morgantown, WV 26506, USA}
\affiliation{Center for Gravitational Waves and Cosmology, West Virginia University, Chestnut Ridge Research Building, Morgantown, WV 26505, USA}

\author[0000-0001-6436-8216]{Bjorn Larsen}
\affiliation{Department of Physics, Yale University, New Haven, CT 06520, USA}

\author[0009-0009-2567-1533]{Patrick O'Neill}
\affiliation{Newcastle University, NE1 7RU, UK}

\author[0000-0002-6428-2620]{Rutger van~Haasteren}
\affiliation{Max-Planck-Institut f{\"u}r Gravitationsphysik (Albert-Einstein-Institut), Callinstra{\ss}e 38, D-30167, Hannover, Germany\\
Leibniz Universit{\"a}t Hannover, D-30167, Hannover, Germany}

\author[0000-0002-8935-9882]{Akash Anumarlapudi}
\affiliation{Center for Gravitation, Cosmology and Astrophysics, Department of Physics, University of Wisconsin-Milwaukee,\\ P.O. Box 413, Milwaukee, WI 53201, USA}

\author[0000-0002-4576-9337]{Matteo Bachetti}
\affiliation{Istituto Nazionale di Astrofisica-Osservatorio Astronomico di Cagliari, via della Scienza 5, I-09047 Selargius (CA), Italy}

\author[0000-0002-7965-3076]{Deven Bhakta}
\affiliation{University of Virginia, Department of Astronomy, P.O. Box 400325, Charlottesville, VA 22904, USA}

\author[0000-0001-5438-540X]{Chloe A. Champagne}
\affiliation{Electrical and Computer Engineering Department, Vanderbilt University, Nashville, TN, USA}

\author[0000-0002-6039-692X]{H. Thankful Cromartie}
\affiliation{National Research Council Research Associate, National Academy of Sciences, Washington, DC 20001, USA resident at Naval Research Laboratory, Washington, DC 20375, USA}

\author[0000-0002-6664-965X]{Paul B. Demorest}
\affiliation{National Radio Astronomy Observatory, 1003 Lopezville Rd., Socorro, NM 87801, USA}

\author[0000-0003-1082-2342]{Ross J. Jennings}
\altaffiliation{NANOGrav Physics Frontiers Center Postdoctoral Fellow}
\affiliation{Department of Physics and Astronomy, West Virginia University, P.O. Box 6315, Morgantown, WV 26506, USA}
\affiliation{Center for Gravitational Waves and Cosmology, West Virginia University, Chestnut Ridge Research Building, Morgantown, WV 26505, USA}

\author[0000-0002-0893-4073]{Matthew Kerr}
\affiliation{Space Science Division, Naval Research Laboratory, Washington, DC 20375-5352, USA}

\author[0000-0003-1241-7615]{Sasha Levina}
\affiliation{William H. Miller III Department of Physics and Astronomy, Johns Hopkins University,\\ 3400 N. Charles Street, Baltimore, Maryland, 21218, USA}

\author[0000-0001-5481-7559]{Alexander McEwen}
\affiliation{Center for Gravitation, Cosmology and Astrophysics, Department of Physics, University of Wisconsin-Milwaukee,\\ P.O. Box 413, Milwaukee, WI 53201, USA}

\author[0000-0002-7283-1124]{Brent J. Shapiro-Albert}
\affiliation{Department of Physics and Astronomy, West Virginia University, P.O. Box 6315, Morgantown, WV 26506, USA}
\affiliation{Center for Gravitational Waves and Cosmology, West Virginia University, Chestnut Ridge Research Building, Morgantown, WV 26505, USA}
\affiliation{Giant Army, 915A 17th Ave, Seattle WA 98122, USA}

\author[0000-0002-1075-3837]{Joseph K. Swiggum}
\altaffiliation{NANOGrav Physics Frontiers Center Postdoctoral Fellow}
\affiliation{Department of Physics, Lafayette College, Easton, PA 18042, USA}

\correspondingauthor{Abhimanyu Susobhanan}
\email{abhimanyu.susobhanan@nanograv.org}




\begin{abstract}
\pint{} is a pure-Python framework for high-precision pulsar timing developed on top of widely used and well-tested Python libraries, supporting both interactive and programmatic data analysis workflows.
We present a new frequentist framework within \pint{} to characterize the single-pulsar noise processes present in pulsar timing datasets.
This framework enables the parameter estimation for both uncorrelated and correlated noise processes as well as the model comparison between different timing and noise models \updated{in a computationally inexpensive way}.
We demonstrate the efficacy of the new framework by applying it to simulated datasets as well as a real dataset of PSR B1855+09.
We also describe the new features implemented in \pint{} since it was first described in the literature.
\end{abstract}



\section{Introduction} 
\label{sec:intro}

Since their discovery, pulsars have been used as celestial laboratories to probe a wide range of time-domain phenomena owing to their remarkable rotational stability.
This is especially true for millisecond pulsars (MSPs), which are pulsars with millisecond-scale rotational periods spun up by accretion from \updated{their companion stars} \citep{Manchester2017}.
Such applications include constraints on the neutron star equation of state \citep[e.g.][]{CromartieFonseca+2020}, discovery of exoplanets \citep{WolszczanFrail1992}, tests of theories of gravity \citep[e.g.][]{KramerStairs+2021}, probing the interstellar medium \citep[e.g.][]{DonnerVerbiest+2020} and solar wind \citep[e.g.][]{TiburziShaifullah+2021}, creation of an international time standard \citep{HobbsGuo+2019}, characterizing the uncertainties present in the solar system ephemerides \citep[e.g.][]{CaballeroGuo+2018}, and more.
These exciting results were produced with the help of pulsar timing, the technique of tracking a pulsar's rotational phase using the measured times of arrival (TOAs) of its pulses, allowing it to be used as a celestial clock \citep{LorimerKramer2012}.
Pulsar timing was instrumental in the recent evidence for a nanohertz gravitational wave background \citep{AgazieAnumarlapudi+2023_ng15gwb, AntoniadisArumugam+2023_eptadr2gwb, ReardonZic+2023, XuChen+2023, AgazieAntoniadis+2024} by Pulsar Timing Array (PTA) experiments \citep{Sazhin1978, FosterBacker1990}, inaugurating the era of nanohertz gravitational wave astronomy.

High-precision pulsar timing experiments such as PTAs and tests of gravity require modeling the TOAs down to nanosecond-level precision.
A pulsar timing model or pulsar ephemeris is a generative mathematical description of the deterministic astrophysical processes influencing the measured TOAs.
These processes include pulsar rotation, pulsar binary dynamics, interstellar dispersion, solar system dynamics, proper motion, solar wind, etc., and must be accurately incorporated into the timing model to achieve the required precision \citep{EdwardsHobbsManchester2006}.
The timing model is often accompanied by a noise model, which incorporates stochastic processes affecting the TOAs such as radiometer noise, pulse jitter, rotational irregularities, interstellar medium variability, radio frequency interference (RFI), etc \citep{AgazieAnumarlapudi+2023_ng15detchar}.

In practice, pulsar timing involves the creation and incremental refinement of a pulsar timing model that matches the observed TOAs, typically using frequentist methods.
This is usually performed using one of the three standard software packages: \tempo{} \citep{NiceDemorest+2015}, \tempotwo{} \citep{HobbsEdwardsManchester2006,EdwardsHobbsManchester2006}, and \pint{} \citep{LuoRansom+2021}, often in an interactive manner.
Noise characterization is usually performed separately in the Bayesian paradigm using software packages such as \enterprise{} \citep{JohnsonMeyers+2024} and \temponest{} \citep{LentatiHobson+2014}, starting from a post-fit timing model.
\enterprise{} can also be used to characterize deterministic and stochastic signals common across multiple pulsars, such as the stochastic gravitational wave background and solar system ephemeris errors \citep[e.g.][]{AgazieAnumarlapudi+2023_ng15gwb, VallisneriTaylor+2020}.
\updated{Since accurate noise characterization is necessary for accurate timing and vice versa \citep[e.g.][]{ColesHobbs+2011}, the timing and noise characterization steps must be repeated alternately to ensure that the results remain accurate and stable, and this turns out to be expensive both in time and computation.}

\pint{}\footnote{Available as \texttt{pint-pulsar} via \texttt{pip} and \texttt{conda} package managers. 
The source code is available at \url{https://github.com/nanograv/PINT}. The documentation is available at \url{https://nanograv-pint.readthedocs.io/}. 
\updated{This paper corresponds to \pint{} v1.0.0 (\url{https://zenodo.org/records/11396241}).
`\pint{}' is an acronym for `\pint{} Is Not \texttt{TEMPO3}'.}} is a flexible pure-Python framework for pulsar timing that is written on top of widely-used scientific computing libraries such as \texttt{numpy} \citep{HarrisMillman+2020}, \texttt{scipy} \citep{VirtanenGommers+2020},  \texttt{astropy} \citep{Price-WhelanLim+2022}, and \texttt{matplotlib} \citep{Hunter2007}, developed under the aegis of the North American Nanohertz Observatory for Gravitational Waves \citep[NANOGrav:][]{DemorestFerdman+2013}.
The reliability of this package is ensured via its reliance on these well-tested libraries, strict version control, and an extensive continuous integration and testing suite.
\pint{} is primarily designed to be used as a Python library to ensure that (a) \emph{all} of its functionality remains easily accessible to the user, (b) it is easily extensible, and (c) it can be easily composed with other Python packages. 
It also provides a graphical user interface (named \texttt{pintk}) and command-line tools for specific tasks.
In comparison, \tempo{} and \tempotwo{}, written in FORTRAN and  C-style C++ respectively, are primarily designed to be used as command-line applications (\tempotwo{} also has a graphical user interface named \texttt{plk} and a Python wrapper named \texttt{libstempo}; \citealt{Vallisneri2020}).

In this \updated{paper}, we present a new frequentist framework in \pint{} for characterizing the noise processes affecting pulsar timing, allowing the noise parameters to be fit \updated{together} with the timing model parameters in a maximum-likelihood way for a single pulsar.
\updated{Our} framework also enables model comparisons within \pint{} using the Akaike Information Criterion \citep[AIC:][]{BurnhamAnderson2004}.
The new framework should allow us to easily incorporate noise characterization into \updated{pulsar timing pipelines and interactive workflows}, obtaining relatively quick noise estimates, \updated{and enabling swift refinement of noise models}.
This is in contrast to conventional Bayesian noise characterization, which is performed as a separate step from pulsar timing and is relatively more computationally expensive but can also include common noise terms between pulsars \citep[e.g.][]{AgazieAnumarlapudi+2023_ng15detchar}.
\updated{Our framework can complement the Bayesian methods for noise characterization in the following ways.
During the initial data preparation and combination stages, our method can hasten the iterative refinement of noise models.
The frequentist estimates can be used to cross-check the Bayesian results.} 
They can also act as starting points for Markov Chain Monte Carlo (MCMC) samplers \citep[e.g.][]{JonesQin2022} allowing them to burn in faster. 
\updated{Finally, in cases where Bayesian analysis is not considered worth the cost, our method provides an inexpensive alternative for noise characterization.}

This paper is arranged as follows.
Section \ref{sec:pulsar-timing-basics} provides a quick overview of pulsar timing.
Section \ref{sec:timingfit} briefly describes the fitting methods used in \pint{}.
The newly implemented methods for estimating noise parameters are described in Section \ref{sec:noisefit}.
Model comparison using the AIC is described in Section \ref{sec:modelcomp}.
We demonstrate the new framework using simulations in Section \ref{sec:sims}, and using a real dataset in Section \ref{sec:b1855+09}.
In Section \ref{sec:newdevs}, we discuss some of the new developments in \pint{}, implemented since the publication of \citet{LuoRansom+2021} which initially described the package.
Finally, we summarize our work in Section \ref{sec:summary}.

\section{A brief overview of pulsar timing} 
\label{sec:pulsar-timing-basics}

\subsection{TOAs}

The primary measurable quantity in pulsar timing is the TOA.
In conventional pulsar timing, the pulsar time series data are coherently averaged (`folded') into an integrated pulse profile to improve its signal-to-noise ratio (S/N) and to mitigate the effects of pulse-to-pulse variations \citep{LorimerKramer2012}.
A TOA can be measured from an integrated pulse profile by matching it against a noise-free template \citep{Taylor1992}.
In the traditional narrowband paradigm, the observation is split into multiple frequency sub-bands, and the TOAs are estimated in each sub-band independently.
On the other hand, in the more recently developed wideband paradigm, a frequency-resolved integrated pulse profile is cross-correlated against a 2-dimensional template in frequency and pulse phase to simultaneously measure a TOA and a dispersion measure (DM)\footnote{DM quantifies the interstellar dispersion of the radio waves and is proportional to the electron column density along the line of sight to the pulsar.} for the whole observation \citep{PennucciDemorestRansom2014, Pennucci2019}.
Algorithms for folding and manipulating integrated pulse profiles and for measuring TOAs are available in packages like \texttt{DSPSR} \citep{vanStratenBailes2011}, \texttt{PRESTO} \citep{Ransom2001},  \texttt{PSRCHIVE} \citep{HotanVanStratenManchester2004}, and \texttt{PulsePortraiture} \citep{PennucciDemorestRansom2014, Pennucci2019}.
We restrict ourselves to the narrowband paradigm in this work for the sake of simplicity unless explicitly stated otherwise.

The TOAs are generally recorded against local observatory clocks.
\pint{} applies a series of clock corrections to the measured TOAs, bringing them to the Barycentric Dynamical Time (TDB), a relativistic timescale defined at the solar system barycenter (SSB).
Detailed descriptions of clock corrections may be found in \citet{HobbsEdwardsManchester2006} and \citet{LuoRansom+2021}.
Note that \tempotwo{} uses the Barycentric Coordinate Time (TCB) by default, which differs from TDB by a constant factor, and timing models using TCB need to be converted to TDB to be \pint{}-compatible (see Subsection \ref{sec:convutils} for the newly-available TCB to TDB conversion feature).

In \pint{}, a set of observed TOAs are represented by the \texttt{TOAs} class in the \texttt{pint.toa} module.
See \citet{LuoRansom+2021} for details on the internal representation of TOAs.
They are usually stored in human-readable text files known as `tim' files
\updated{which} can be read using the \texttt{pint.toa.get\_TOAs()} function.

\subsection{The timing and noise model}
\label{sec:timing-model}

Pulsar timing involves connecting the pulse number $N$, related to the rotational phase of the pulsar as $\Phi(N)=2\pi N$, to the time of emission $t_\text{em}$ as 
\begin{equation}
    N = N_0 + f(t_\text{em} - t_0) + \frac{1}{2}\dot{f}(t_\text{em} - t_0)^2 + ... \,,
    \label{eq:phase}
\end{equation}
where $f$ is the rotational frequency, $\dot{f}$ is the rotational frequency derivative, and $t_0$ is a fiducial time.
The right-hand side of the above equation may also include higher-order frequency derivative terms and rotational irregularity effects such as glitches \citep{HobbsEdwardsManchester2006}.
The measured TOA $t_\text{arr}$ is related to $t_\text{em}$ by 
\begin{align}
    t_\text{arr} = t_\text{em} + \Delta_\text{B} + \Delta_\text{DM} + \Delta_\text{\astrosun} + ... + \mathcal{N}\,.
    \label{eq:delays}
\end{align}
Here, the delay $\Delta_\text{B}$ originates from the motion of the pulsar in a binary system and includes R{\o}mer delay, Shapiro delay, and Einstein delay \citep{DamourDeruelle1986}.
$\Delta_\text{DM}$ denotes the delay caused by interstellar dispersion and is given by $\Delta_\text{DM}=\mathcal{K} D/\nu^2$, where $\nu$ is the observing frequency, $D$ is the DM, and $\mathcal{K}$ is known as the DM constant \citep{LorimerKramer2012}.
$\Delta_\text{\astrosun}$ denotes the delays caused by the Solar System motion, including the R{\o}mer delay and the Shapiro delay\footnote{\updated{Solar system Einstein delay is corrected for while converting the TOAs into TDB, since it is part of the definition of the TDB.}}, and are computed using the solar system ephemerides published by space agencies \citep[e.g.,][]{ParkFolkner+2021}. 
\updated{Detailed descriptions} of the various timing model components can be found in \citet{EdwardsHobbsManchester2006} and \citet{LuoRansom+2021}.
Finally, $\mathcal{N}$ denotes the noise present in the TOA, including correlated and uncorrelated noise components (see Subsection \ref{sec:noisetypes}).
\updated{Note that, in equation \eqref{eq:delays}, we have ignored an un-measurable constant term corresponding to the light travel time in vacuum from the pulsar (or the binary barycenter in the case of binary pulsars) to the SSB at some fiducial epoch.}

The various timing \updated{and noise} model components available in \pint{} are listed in Table \ref{tab:model-components}, and the new/updated components are highlighted therein.
These are available in the \texttt{pint.models} module of \pint{}.
The timing \updated{and noise} model as a whole, comprising these components, is represented by the \texttt{TimingModel} class in the \texttt{pint.models} module.
Pulsar ephemerides are usually stored in human-readable text files known as `par' files and can be read using the \texttt{pint.models.get\_model()} function. 
A pair of `par' and `tim' files belonging to the same pulsar can be read together using the \texttt{pint.models.get\_model\_and\_toas()} function.\footnote{The latter is the preferred way since some parts of the timing model, such as clock and solar system ephemeris information, can affect how the \texttt{TOAs} object is constructed.}



\begin{center}
\begin{longtable*}{c|l|c}
\hline 
\textbf{Component} & \textbf{Description} & \textbf{References}\tabularnewline
\hline 
\hline 
\multicolumn{3}{l}{\textbf{\emph{Pulsar rotation, rotational phase \& rotational irregularities}}}\tabularnewline
\hline 
\texttt{Spindown} & Taylor series representation of the pulsar rotation & \citet{BackerHellings1986} \tabularnewline
\arrayrulecolor{gray}\hline
\texttt{PiecewiseSpindown \S} & Piecewise-constant corrections to pulsar rotation & \tabularnewline
\hline 
\texttt{Glitch} & Pulsar glitches & \citet{HobbsEdwardsManchester2006} \tabularnewline
\hline 
\texttt{IFunc} & Piecewise-constant or spline representation of rotational period & \citet{DengColes+2012}\tabularnewline
\hline 
\texttt{WaveX \S \ddag} & Fourier series representation of achromatic red noise (ARN) & \citet{HobbsEdwardsManchester2006} \tabularnewline
 & (supersedes the deprecated \texttt{Wave} model) & Subsection \ref{sec:corrnoise}, \ref{sec:rednoisefit}  \tabularnewline
\hline 
\texttt{PLRedNoise} & Fourier Gaussian process representation of achromatic red & \citet{LentatiHobson+2014}\tabularnewline
 & noise & \tabularnewline
\hline 
\texttt{AbsPhase} & Reference TOA with respect to which the rotational phase  & \tabularnewline
 & is measured & \tabularnewline
\hline 
\texttt{PhaseOffset \S \ddag} & Overall phase offset between reference TOA and the physical & Subsection \ref{sec:whitenoisefit} \tabularnewline
 & TOAs & \tabularnewline
\hline 
\texttt{PhaseJump \P} & Phase offsets between TOAs measured using different systems & \citet{HobbsEdwardsManchester2006} \tabularnewline
\arrayrulecolor{black}
\hline 
\multicolumn{3}{l}{\textbf{\emph{Binary system}}}\tabularnewline
\hline 
\arrayrulecolor{gray}
\texttt{BinaryBT} & Simple parametrized post-Keplerian binary model  & \citet{BlandfordTeukolsky1976} \tabularnewline
 & (only R{\o}mer delay) & \tabularnewline
\hline 
\texttt{BinaryBTPiecewise \S \ddag} & Similar to BinaryBT, but with piecewise-constant & \tabularnewline
 & orbital parameters & \tabularnewline
\hline 
\texttt{BinaryDD} & Parametrized post-Keplerian model (with Shapiro & \citet{DamourDeruelle1986}\tabularnewline
 & delay and Einstein delay) & \tabularnewline
\hline 
\texttt{BinaryDDGR \S} & Similar to \texttt{BinaryDD}, but assumes General Relativity & \citet{TaylorWeisberg1989} \tabularnewline
\hline 
\texttt{BinaryDDH \S} & Similar to \texttt{BinaryDD}, but uses a harmonic representation & \citet{FreireWex2010} \tabularnewline
 & of Shapiro delay (for low-inclination systems) & \citet{WeisbergHuang2016} \tabularnewline
\hline 
\texttt{BinaryDDK} & Similar to \texttt{BinaryDD}, but includes Kopeikin delay & \citet{DamourTaylor1992} \tabularnewline
 &  & \citet{Kopeikin1995,Kopeikin1996} \tabularnewline
\hline 
\texttt{BinaryDDS \S} & Similar to \texttt{BinaryDD}, but uses an alternative representation & \citet{KramerStairs+2006} \tabularnewline
 & of Shapiro delay (for almost edge-on orbits) & \citet{RafikovLai2006}
 \tabularnewline
\hline 
\texttt{BinaryELL1 \dag} & Binary model specialized for nearly circular orbits using & \citet{LangeCamilo+2001} \tabularnewline
 & Laplace-Lagrange parameters (includes up to third-order & \citet{ZhuDesvignes+2018} \tabularnewline
 & terms in eccentricity)  & \citet{FioreLevin+2023} \tabularnewline
\hline 
\texttt{BinaryELL1H} & Similar to \texttt{BinaryELL1}, but uses a harmonic representation  & \citet{FreireWex2010} \tabularnewline
 & of Shapiro delay (for low-inclination systems) & \tabularnewline
\hline 
\texttt{BinaryELL1k \S} & Similar to \texttt{BinaryELL1}, but includes an exact treatment of  & \citet{SusobhananGopakumar+2018} \tabularnewline
 & advance of periastron (for highly relativistic or tidally   & \tabularnewline
 & interacting binaries) & \tabularnewline
\arrayrulecolor{black}
\hline 
\multicolumn{3}{l}{\textbf{\emph{Interstellar dispersion \& dispersion measure variations}}}\tabularnewline
\hline 
\arrayrulecolor{gray}
\texttt{DispersionDM \P} & Taylor series representation of dispersion measure (DM) & \citet{BackerHellings1986} \tabularnewline
\hline 
\texttt{DispersionDMX \P} & Piecewise-constant representation of DM variations & \citet{ArzoumanianBrazier+2015} \tabularnewline
\hline 
\texttt{DMWaveX \S \ddag} & Fourier series representation of DM variations & Subsection \ref{sec:corrnoise}, \ref{sec:rednoisefit} \tabularnewline 
\hline 
\texttt{PLDMNoise \S} & Fourier Gaussian process representation of DM variations & \citet{LentatiHobson+2014} \tabularnewline
\hline 
\texttt{DispersionJump \S} & Offsets between wideband DMs measured using different & \citet{AlamArzoumanian+2021} \tabularnewline
 & systems (no delay) & \tabularnewline
 \hline 
\texttt{FDJumpDM \S} & DM offsets between narrowband TOAs measured using  & 
\tabularnewline
 & different systems & \tabularnewline
\arrayrulecolor{black}
\hline 
\multicolumn{3}{l}{\textbf{\emph{Astrometry \& solar system delays}}}\tabularnewline
\hline 
\arrayrulecolor{gray}
\texttt{AstrometryEcliptic} & Astrometry in ecliptic coordinates & \citet{EdwardsHobbsManchester2006} \tabularnewline
\cline{1-2} \cline{2-2} 
\texttt{AstrometryEquatorial} & Astrometry in equatorial coordinates & \tabularnewline
\hline 
\texttt{SolarSystemShapiro} & Solar system Shapiro delay & \citet{Shapiro1964} \tabularnewline
\arrayrulecolor{black}
\hline 
\multicolumn{3}{l}{\textbf{\emph{Solar wind}}}\tabularnewline
\hline 
\arrayrulecolor{gray}
\texttt{SolarWindDispersion \dag} & Solar wind model assuming a radial power-law relation for & \citet{EdwardsHobbsManchester2006} \tabularnewline
 & the electron density & \citet{YouHobbs+2007,YouColes+2012}\tabularnewline
\cline{1-2} \cline{2-2} 
\texttt{SolarWindDispersionX \S \ddag} & Similar to SolarWindDispersion, but with a piecewise- & \citet{MadisonCordes+2019} \tabularnewline
 & constant representation of the electron density. & \citet{HazbounSimon+2022} \tabularnewline
 \arrayrulecolor{black}
\hline 
\multicolumn{3}{l}{\textbf{\emph{Troposphere}}}\tabularnewline
\hline 
\texttt{TroposphereDelay} & Tropospheric zenith hydrostatic delay & \citet{DavisHerring+1985} \tabularnewline
 &  & \citet{Niell1996} \tabularnewline
\hline 
\multicolumn{3}{l}{\textbf{\emph{Time-uncorrelated noise}}}\tabularnewline
\hline 
\arrayrulecolor{gray}
\texttt{ScaleToaError} & Modifications to the measured TOA uncertainties   & \citet{LentatiHobson+2014} \tabularnewline
\hline 
\texttt{ScaleDMError \S} & Modifications to the measured wideband DM uncertainties & \citet{AlamArzoumanian+2021} \tabularnewline
\hline 
\texttt{EcorrNoise \dag} & Correlation between TOAs measured from the same  & \citet{ArzoumanianBrazier+2015} \tabularnewline
 & observation.  & \citet{JohnsonMeyers+2024} \tabularnewline
 \arrayrulecolor{black}
\hline 
\multicolumn{3}{l}{\textbf{\emph{Frequency-dependent profile evolution}}}\tabularnewline
\hline 
\arrayrulecolor{gray}
\texttt{FD \P} & Frequency-dependent profile evolution & \citet{ArzoumanianBrazier+2015} \tabularnewline
\hline 
\texttt{FDJump \S} & System and frequency-dependent profile evolution & Appendix \ref{sec:fdjump} \tabularnewline
\arrayrulecolor{black}
\hline 
\caption{Updated list of timing model components available in \pint{} (in the \texttt{pint.models} module).
`\S' denotes newly implemented components.
`\dag' denotes components that have had significant changes since the publication of \citet{LuoRansom+2021}.
The \texttt{BinaryELL1} component now includes up to $\mathcal{O}(e^3)$ terms in orbital eccentricity.
The \texttt{SolarWindDispersion} component now allows electron density radial power law indices other than 2.
The \texttt{EcorrNoise} component now incorporates a faster algorithm for inverting the TOA covariance matrix.
`\P' denotes components that were renamed after the publication of \citet{LuoRansom+2021}.
`\ddag' denotes components only available in \pint{}.\\
The choice of the binary model is specified in the `par' files using the `\texttt{BINARY}' keyword, e.g., `\texttt{BINARY ELL1}'. 
Additionally, there is a generic timing model in \tempotwo{} called the `\texttt{T2}' model, which chooses the underlying binary model based on the available binary parameters.
When `\texttt{BINARY T2}' is encountered, \pint{} emits an informative error message indicating the best guess for the underlying binary model. 
\updated{It can also construct the binary timing model based on this guess optionally.}}
\label{tab:model-components}
\end{longtable*}
\end{center}

\newpage

\subsection{Timing residuals}

The timing model can be used to predict the phase $\Phi_i$ associated with each TOA, and this allows us to compute the timing residuals
\begin{align}
    r_i = \frac{\Phi_i - 2\pi N[\Phi_i]}{F}\,,
    \label{eq:residual}
\end{align}
where $N[\Phi_i]$ is the integer closest to $\Phi_i/2\pi$, and \updated{$F$ is the pulse frequency}\footnote{\updated{In \pint{}, $F$ can be the rotational frequency $F_0$ at some fiducial epoch, the instantaneous rotational frequency $F(t_i)$ computed incorporating the rotational frequency derivatives, or the topocentric pulse frequency $\bar{F}(t_i)=d\Phi/dt_{\text{arr}}$ incorporating all delay and phase corrections computed using finite differences. 
On the other hand, \tempotwo{} uses $F=F_0$, and \tempo{} uses the instantaneous rotational frequency that also incorporates the first rotational frequency derivative.
The topocentric frequency is the most accurate option in principle, but it is more expensive to compute and the difference between different options is negligible for MSPs in most cases. 
However, not using the topocentric frequency can have a significant impact on the timing accuracy in the case of young or accreting pulsars with large frequency derivatives \cite[e.g.][]{RayGuillot+2019}.}}.
The procedure for computing a timing residual can be summarized as follows:
\begin{enumerate}
    \item Apply the various clock corrections to convert the measured TOA from the observatory timescale to a \updated{timescale defined at the SSB}.
    \item Successively correct for the various delays influencing the TOA, bringing it to the pulsar frame (equation \ref{eq:delays}).
    \item Compute the pulsar rotational phase at the corrected TOA using the rotational frequency and its derivatives (equation \ref{eq:phase}).
    \item Successively correct for the other effects that influence the rotational phase of the pulsar.
    \item Compute the phase residual by subtracting the closest integer from the estimated rotational phase. The timing residual is the phase residual divided by the instantaneous rotational frequency (equation \ref{eq:residual}).
\end{enumerate}

In the \texttt{pint.residuals} module, narrowband timing residuals are represented by the \texttt{Residuals} class, and wideband residuals are represented by the \texttt{WidebandTOAResiduals} class.
\vspace{0.1in}

How we fit a timing model to the observed TOAs using timing residuals is described in the next Section.

\section{Fitting for timing model parameters}
\label{sec:timingfit}

\subsection{Fitting in the white noise-only case}

If the initial (pre-fit) timing model is sufficiently close to its best-fit counterpart, the pre-fit timing residuals $\boldsymbol{r}$ and the post-fit timing residuals $\boldsymbol{s}$ are related by the linear relation
\begin{align}
    \boldsymbol{r} - \boldsymbol{s} = \boldsymbol{M}\boldsymbol{\beta}\,,
    \label{eq:linear-tm}
\end{align}
where $\boldsymbol{r}$ and $\boldsymbol{s}$ are $n$-dimensional vectors containing elements $r_i$ and $s_i$ respectively, $\boldsymbol{M}$ is the $n\times p$-dimensional pulsar timing design matrix containing partial derivatives $\frac{\partial s_i}{\partial b_\alpha}$ with respect to the $p$ timing model parameters $b_\alpha$, and $\boldsymbol{\beta}$ is a $p$-dimensional vector containing timing model parameter deviations $\beta_\alpha$ from their best-fit values $\hat{b}_\alpha$ (i.e., $\beta_\alpha=\hat{b}_\alpha-b_\alpha$), $n$ is the number of TOAs, and $p$ is the number of timing model parameters.
In the absence of correlated noise, the log-likelihood function can be written, up to an additive constant, as
\begin{align}
    \ln L = -\frac{1}{2} \boldsymbol{s}^T \boldsymbol{N}^{-1} \boldsymbol{s} - \frac{1}{2} \ln\det \boldsymbol{N}\,,
\end{align}
where $\boldsymbol{N}=\text{diag}[\varsigma_i^2]$ is the diagonal uncorrelated (white) noise TOA covariance matrix, and $\varsigma_i$ represent the scaled TOA uncertainties (see Section \ref{sec:noisefit} for a detailed explanation). 
The quantity $\boldsymbol{s}^T \boldsymbol{N}^{-1} \boldsymbol{s}$ appearing in the first term is usually referred to as the chi-squared ($\chi^2$).

The parameter deviations $\boldsymbol{\beta}$ can be estimated by maximizing the above likelihood function.
If $\boldsymbol{N}$ is fixed, the maximum-likelihood estimate involves minimizing the $\chi^2$, and this turns out to be
\begin{align}
    \hat{\boldsymbol{\beta}} = \left(\boldsymbol{M}^T\boldsymbol{N}^{-1}\boldsymbol{M}\right)^{-1} \boldsymbol{M}^T \boldsymbol{N}^{-1} \boldsymbol{r}\,,
    \label{eq:betahat_wls}
\end{align}
with a parameter covariance matrix
\begin{align}
    \boldsymbol{K}_\beta = \frac{1}{n-p} (\boldsymbol{s}^T \boldsymbol{N}^{-1} \boldsymbol{s}) (\boldsymbol{M}^T \boldsymbol{N}^{-1} \boldsymbol{M})^{-1}\,.
\end{align}
The conventional fitting algorithm for the pulsar timing model involves updating the parameter values $\boldsymbol{b} \rightarrow \boldsymbol{b} - \hat{\boldsymbol{\beta}}$. 
\updated{In practice, the R.H.S. of equation \eqref{eq:betahat_wls} is evaluated with the help of a singular value decomposition (SVD) in \pint{} (see also Subsection \ref{sec:param-degen}) \citep{PressTeukolsky+1992}.
In contrast, \tempotwo{} uses a QR decomposition for this purpose.}

This method is usually referred to as weighted least-squares (WLS).
\updated{See, e.g., \citet{ColesHobbs+2011} for a detailed discussion on this method.}

\subsection{Fitting in the presence of correlated noise}

The more general case of the above fitting algorithm that accounts for the presence of correlated noise involves maximizing the likelihood function 
\begin{align}
    \ln L = -\frac{1}{2} \boldsymbol{s}^T \boldsymbol{C}^{-1} \boldsymbol{s} - \frac{1}{2} \ln\det \boldsymbol{C}\,,
    \label{eq:gls-like}
\end{align}
where $\boldsymbol{C}$ is the non-diagonal covariance matrix that incorporates both white and correlated noise. 
Given fixed noise parameters (i.e., fixed $\boldsymbol{C}$), the maximum-likelihood values for the parameter deviations can be written formally as \citep{ColesHobbs+2011}
\begin{align}
    \hat{\boldsymbol{\beta}} = \left(\boldsymbol{M}^T\boldsymbol{C}^{-1}\boldsymbol{M}\right)^{-1} \boldsymbol{M}^T \boldsymbol{C}^{-1} \boldsymbol{r}\,,
    \label{eq:betahat}
\end{align}
along with the parameter covariance matrix
\begin{align}
    \boldsymbol{K}_\beta = \frac{1}{n-p} (\boldsymbol{s}^T \boldsymbol{C}^{-1} \boldsymbol{s}) (\boldsymbol{M}^T \boldsymbol{C}^{-1} \boldsymbol{M})^{-1}\,.
    \label{eq:Kbeta}
\end{align}
Once $\hat{\boldsymbol{\beta}}$ are computed, the parameter values can be updated as $\boldsymbol{b} \rightarrow \boldsymbol{b} - \hat{\boldsymbol{\beta}}$ similar to the WLS case, and this method is usually referred to as the generalized least-squares (GLS). 
\updated{Note that equation \eqref{eq:betahat} is evaluated in \pint{} using a Cholesky decomposition for non-degenerate systems and using an SVD for degenerate systems (see Subsection \ref{sec:param-degen} for details)}. 
 
Since $\boldsymbol{C}$ is in general not diagonal, computing $\boldsymbol{C}^{-1}$ is much more expensive than computing $\boldsymbol{N}^{-1}$, and scales as $\mathcal{O}(n^3)$ as opposed to $\mathcal{O}(n)$ in the worst case scenario. 
Fortunately, $\boldsymbol{C}$ can be represented in most cases as a rank-$m$ update to the diagonal matrix $\boldsymbol{N}$ as \citep{VanHaasterenVallisneri2014a}
\begin{align}
    \boldsymbol{C} = \boldsymbol{N} + \boldsymbol{U}\boldsymbol{\Phi}\boldsymbol{U}^T\,,
    \label{eq:woodbury}
\end{align}
where $\boldsymbol{U}$ is an $n\times m$-dimensional correlated noise basis matrix and $\boldsymbol{\Phi}$ is a $m\times m$-dimensional diagonal matrix containing the correlated noise weights.
In this case, $\boldsymbol{C}^{-1}$ can be written using the Woodbury identity as 
\begin{align}
    \boldsymbol{C}^{-1} = \boldsymbol{N}^{-1} - \boldsymbol{N}^{-1}\boldsymbol{U}\boldsymbol{\Sigma}^{-1}\boldsymbol{U}^T\boldsymbol{N}^{-1}\,,
\end{align}
where
\begin{align}
    \boldsymbol{\Sigma}^{-1} = \boldsymbol{\Phi}^{-1} + \boldsymbol{U}^T\boldsymbol{N}^{-1}\boldsymbol{U}\,.
\end{align}
Additionally, $\det\boldsymbol{C}$ appearing in the $\ln L$ expression can be computed using the identity
\begin{align}
    \det \boldsymbol{C} = 
    \det \boldsymbol{N} \times 
    \det \boldsymbol{\Sigma} \times  \det \boldsymbol{\Phi}\,.
\end{align}
This allows us to evaluate $\ln L$, $\hat{\boldsymbol{\beta}}$, and $\boldsymbol{K}_\beta$ with $\mathcal{O}(nm^2)$ time complexity assuming $m \ll n$.

\subsection{Failure modes of linear fitting and the Downhill Fitter algorithm}

\subsubsection{Handling parameter degeneracies}
\label{sec:param-degen}
The above-described fitting algorithm, while successful in the vast majority of cases, can nevertheless fail to correctly estimate the maximum-likelihood parameters under certain conditions.
The most obvious such scenario is parameter degeneracy, which leads to $\boldsymbol{M}^T \boldsymbol{C}^{-1} \boldsymbol{M}$ being singular.
Ideally, this should be addressed by reparametrizing the timing model to avoid the degeneracy.
Alternatively, it can be addressed in an ad hoc manner by restricting the fitting algorithm to operate only in a subspace of the parameter space where the fitting problem is non-singular.
This is done by replacing the inverse $(\boldsymbol{M}^T \boldsymbol{C}^{-1} \boldsymbol{M})^{-1}$ by the pseudoinverse $\boldsymbol{V} \boldsymbol{\bar{S}}^{-1} \boldsymbol{U}^T$, where 
$\boldsymbol{M}^T \boldsymbol{C}^{-1} \boldsymbol{M} = \boldsymbol{U} \boldsymbol{S} \boldsymbol{V}^T$
is an SVD such that $\boldsymbol{U}$ and $\boldsymbol{V}^T$ are orthogonal matrices, $\boldsymbol{S}$ is a diagonal matrix containing the singular values, and ${\bar{\boldsymbol{S}}}^{-1}$ is obtained by replacing the diagonal elements of $\boldsymbol {S}^{-1}$ which are greater than a certain threshold by 0.

\subsubsection{The Downhill Fitter algorithm}
\label{sec:downhill}

Unfortunately, the fitting algorithm can fail even in the absence of parameter degeneracies under the following conditions: (a) the linear approximation underlying equation \eqref{eq:linear-tm} breaks down and the non-linear terms become important, and (b) the $\boldsymbol{b} \rightarrow \boldsymbol{b} - \hat{\boldsymbol{\beta}}$ update brings a parameter outside its physically meaningful range (e.g., orbital eccentricity $e\in[0,1)$, Shapiro delay shape $\sin\iota\in[0,1]$) such that the likelihood function becomes ill-defined.
\pint{} implements a robust fitting algorithm, named the Downhill fitter, to deal with such cases, and it is briefly described below.

In the Downhill fitter, the usual $\boldsymbol{b} \rightarrow \boldsymbol{b} - \hat{\boldsymbol{\beta}}$ update is replaced by a more general update $\boldsymbol{b} \rightarrow \boldsymbol{b} - \lambda \hat{\boldsymbol{\beta}}$, with $\lambda \in (0,1]$. 
Even in the problematic cases mentioned above, it may be possible to find a value of $\lambda$ that leaves the likelihood function both well-defined and improved even if $\lambda=1$ does not provide an acceptable solution.
In practice, this is done by reducing $\lambda$ iteratively by some factor, starting from 1, until an acceptable solution is found.
If an improved solution cannot be found even after a certain maximum number of iterations of reducing $\lambda$, the original solution is retained.

~

The various fitting methods available in \pint{} (in the \texttt{pint.fitter} module) are listed in Table \ref{tab:fitters} and the new/updated features are highlighted therein.  We have also added a method \texttt{pint.fitter.Fitter.auto()} which selects the appropriate type of fitter depending on the input data (narrowband vs.\ wideband, white noise vs.\ correlated noise) with a preference for the downhill fitter variants.

\begin{table*}[]
    \centering
    \begin{tabular}{c|c|c}
       \hline
       Fitter  &  Description & Reference \\ \hline\hline
       \arrayrulecolor{gray}
        \texttt{PowellFitter} & Fitting using the modified Powell algorithm (uses \texttt{scipy})  & \citet{Powell1964} \\\hline
        \texttt{WLSFitter} & Weighted least-squares fitting & \citet{HobbsEdwardsManchester2006} \\  \hline
        \texttt{DownhillWLSFitter} \S  & Downhill weighted least-squares fitting & Subsection \ref{sec:downhill} \\
         & Allows fitting for noise parameters. &  \\\hline
        \texttt{GLSFitter} & Generalized least-squares fitting & \citet{ColesHobbs+2011} \\ \hline
        \texttt{DownhillGLSFitter} \S  & Downhill generalized least-squares fitting.  & Subsection \ref{sec:downhill} \\  
         & Allows fitting for noise parameters.  &  \\\hline
        \texttt{WidebandTOAFitter}   & Generalized least-squares fitting for wideband TOAs &  \cite{AlamArzoumanian+2021} \\\hline
        \texttt{WidebandDownhillFitter} \S  & Downhill generalized least-squares fitting for wideband TOAs & Subsection \ref{sec:downhill} \\\hline
        \texttt{MCMCFitter} \dag  & MCMC optimization (uses \texttt{emcee}) & \citet{Foreman-MackeyHogg+2013} \\ \arrayrulecolor{black}\hline
    \end{tabular}
    \caption{Fitting methods available in \pint{} (in the \texttt{pint.fitter} module).
    `\dag' denotes fitters that have had significant changes since the publication of \citet{LuoRansom+2021}, and `\S' denotes newly implemented fitters.   
    \texttt{MCMCFitter} was updated to use the \texttt{emcee} package, providing parallel processing capabilities.
    }
    \label{tab:fitters}
\end{table*}

\section{Fitting for noise parameters}
\label{sec:noisefit}

The fitting methods discussed in Section \ref{sec:timingfit} assume that the TOA covariance matrix $\boldsymbol{C}$ (or $\boldsymbol{N}$ in the absence of correlated noise) is fixed.
However, the noise characteristics of a given set of TOAs are not usually known a priori and must be determined from the data itself.
In practice, this is handled by ignoring the noise parameters during the initial data preparation stages, and then performing a separate Bayesian noise analysis step on the data once a reasonable (but not necessarily optimal) timing solution is found.
The timing model is then refined by applying the estimated noise parameters, and this process is iterated.

In this Section, we develop a framework within \pint{} to estimate noise parameters \updated{together} with timing model parameters, obviating the need for computationally expensive Bayesian noise analysis iterations during pulsar timing.
Note that the noise estimates obtained using these methods may still need to be refined \textit{after} the data preparation stage using Bayesian methods. 
Still, the convergence of such analyses can be accelerated by providing the initial noise estimates as starting points for MCMC samplers.

\subsection{Types of noise}
\label{sec:noisetypes}

We begin by briefly discussing the different types of noise seen in pulsar timing.
These processes can be divided into three broad categories, as described below.

\subsubsection{Uncorrelated (white) noise}
Uncorrelated noise or white noise refers to the component of the noise that is independent for each TOA.
It may arise from radiometer noise, pulse jitter, radio frequency interference (RFI), polarization miscalibration, etc.
It is characterized by the diagonal matrix $\boldsymbol{N}$, which is populated by the scaled TOA variances $\varsigma_i^2$, given by
\begin{align}
    \varsigma_i^2 = F_i^2\left( \sigma_i^2 + Q_i^2 \right)\,,
\end{align}
where 
\begin{align}
    F_i &= \prod_a f_a^{\mathcal{F}_{ia}}\,,\\
    Q_i^2 &= \sum_a q_a^2 \mathcal{Q}_{ia}\,.
\end{align}
The quantities $f_a$ and $q_a$ are known as EFACs (`error factors') and EQUADs (`errors added in quadrature') respectively.
$\mathcal{F}_{ia}$ and $\mathcal{Q}_{ia}$ represent TOA selection masks which can be 0 or 1 based on some criterion, which may depend on the observing epoch, observing frequency, observing system, etc.
EFACs and EQUADs are implemented in the \texttt{ScaleToaError} component (see Table \ref{tab:model-components}).

\subsubsection{Time-uncorrelated correlated noise}
Time-uncorrelated correlated noise refers to the component of the noise that is correlated amongst the TOAs derived from the same observation, but are uncorrelated otherwise. 
This is usually referred to as ECORR and may arise from pulse jitter, RFI, polarization miscalibration, interstellar scattering, etc. that are correlated across different frequency subbands of the same observation.
Since ECORR is uncorrelated across different observations, it is possible to express its contribution to the TOA covariance matrix as a sparse block-diagonal matrix. 
This allows us to evaluate $\boldsymbol{C}^{-1}$ using the Sherman-Morrison identity (the rank-1 special case of the Woodbury identity), which turns out to be less expensive than the general case.
The TOA covariance matrix contribution arising from ECORR can be expressed as \citep{JohnsonMeyers+2024}
\begin{align}
    \boldsymbol{C}_\text{ECORR} = \sum_{ab} c_a^2 \boldsymbol{v}_{ab}\boldsymbol{v}_{ab}^T\,,
\end{align}
where $c_a$ represents the ECORR parameter for each selection group, the index $b$ represents the various TOA epochs, and the basis vector $\boldsymbol{v}_{ab}$ contains 1s for each TOA belonging to the epoch $b$ and selection group $a$, and 0s otherwise.
ECORRs are implemented in the \texttt{EcorrNoise} component (see Table \ref{tab:model-components}).

\subsubsection{Time-correlated noise}
\label{sec:corrnoise}
Time-correlated noise in pulsar timing may arise due to the rotational irregularities of the pulsar (spin noise/achromatic red noise, \updated{ARN}), time-variable interstellar dispersion (DM noise, \updated{DMN}), time-variable interstellar scattering (scattering noise), long-timescale RFI (system/band noise), etc.
This type of noise can be represented either as a deterministic signal (such as a piecewise-constant function, Fourier expansion, etc.) or as a Gaussian process (GP) that contributes a dense component to the TOA covariance matrix, usually represented using a reduced rank approximation (see equation \ref{eq:woodbury}) \citep{VanHaasterenVallisneri2014b}.
The deterministic signal representation provides a delay of the form $\boldsymbol{U}\boldsymbol{\varphi}$, where $\boldsymbol{U}$ is a basis matrix and $\boldsymbol{\varphi}$ is an amplitude vector.
For example, the Fourier basis matrix corresponding to a time-correlated noise component is given by
\begin{align}
U_{ij}=\left(\frac{\nu_{\text{ref}}}{\nu_i}\right)^{\alpha}\begin{cases}
\sin\left(\frac{\pi(j+1)(t_{i}-t_{\text{ref}})}{T_\text{span}}\right) & \text{odd }j\\
\cos\left(\frac{\pi j(t_{i}-t_{\text{ref}})}{T_\text{span}}\right) & \text{even }j
\end{cases}\,,
\label{eq:fourier-basis}
\end{align}
where $\nu_i$ is the observing frequency, $\nu_{\text{ref}}$ is an arbitrary reference frequency (conventionally taken as 1400 MHz), $\alpha$ is known as the chromatic index ($\alpha=0$ for \updated{the ARN} and $\alpha=2$ for \updated{the DMN}), $t_i$ is the TOA, $t_{\text{ref}}$ is an arbitrary reference time, and $T_\text{span}$ is the total time span of the TOAs.
The deterministic representation can be converted into a GP representation by analytically marginalizing the amplitudes $\boldsymbol{\varphi}$ assuming Gaussian priors \citep{VanHaasterenVallisneri2014b}. 
The deterministic Fourier representation of the \updated{ARN} is implemented as the \texttt{WaveX} component and that of the \updated{DMN} is implemented as the \texttt{DMWaveX} component.
Their GP counterparts assuming a power-law spectral density are \texttt{PLRedNoise} and \texttt{PLDMNoise} respectively (see Table \ref{tab:model-components}).

\subsection{Fitting for white noise parameters and ECORRs}
\label{sec:whitenoisefit}
If the time-correlated noise components are treated using a deterministic representation as discussed above, the covariance matrix can be written as 
\begin{align}
   \boldsymbol{C}=\boldsymbol{N} + \sum_{ab}c_a^2 \boldsymbol{v}_{ab}\boldsymbol{v}_{ab}^T\,,
\end{align}
which turns out to be block-diagonal. 
Each block $\boldsymbol{C}_{ab}$ can be inverted using the Sherman-Morrison identity \citep{JohnsonMeyers+2024}
\begin{align}\boldsymbol{C}_{ab}^{-1}
&=\left(\boldsymbol{N}_{ab}+c_{a}^{2}\boldsymbol{v}_{ab}\boldsymbol{v}_{ab}^{T}\right)^{-1}\nonumber\\
&=\boldsymbol{N}_{ab}^{-1}-\frac{c_{a}^{2}\boldsymbol{N}_{ab}^{-1}\boldsymbol{v}_{ab}\boldsymbol{v}_{ab}^{T}\boldsymbol{N}_{ab}^{-1}}{1+\boldsymbol{v}_{ab}^{T}\boldsymbol{N}_{ab}^{-1}\boldsymbol{v}_{ab}}\,,
\end{align}
where $\boldsymbol{N}_{ab}$ is the portion of $\boldsymbol{N}$ corresponding to the selection group $a$ and observing epoch $b$.
Similarly, the determinant of each block can be computed using the identity
\begin{align}\det\boldsymbol{C}_{ab}=\det\boldsymbol{N}_{ab}\times\left(1+\boldsymbol{v}_{ab}^{T}\boldsymbol{N}_{ab}^{-1}\boldsymbol{v}_{ab}\right)\,.
\end{align}
Defining the inner product $\left(\boldsymbol{x}|\boldsymbol{y}\right)=\boldsymbol{x}^T \boldsymbol{N}^{-1}\boldsymbol{y}$, the log-likelihood can be expressed as $\ln L = \sum_{ab}\ln L_{ab}$ where
\begin{align}
\ln L_{ab}&=-\frac{1}{2}\Bigg\{\left(\boldsymbol{s}_{ab}|\boldsymbol{s}_{ab}\right)+\ln\det\boldsymbol{N}_{ab}
-\frac{c_{a}^{2}\left(\boldsymbol{s}_{ab}|\boldsymbol{v}_{ab}\right)^{2}}{1+\left(\boldsymbol{v}_{ab}|\boldsymbol{v}_{ab}\right)}
+\ln\left[1+\left(\boldsymbol{v}_{ab}|\boldsymbol{v}_{ab}\right)\right]\Bigg\}\,.
\label{eq:lnL_sm}
\end{align}
We estimate the white noise and ECORR parameters by numerically maximizing the above expression over the parameters $f_a$, $q_a$, and $c_a$ while fixing the timing model parameters.
In practice, we alternate the above maximization procedure with the timing model parameter fitting described in Section \ref{sec:timingfit} several times.
The covariance matrix of the noise parameters $\beta_i\in \left\{f_a, q_a, c_a\right\}$ can be computed by inverting the Hessian matrix $\boldsymbol{H}$ of the log-likelihood function, whose elements are $H_{ij}=\frac{\partial^2 \ln L}{\partial\beta_i \partial\beta_j}$
(the Hessian is computed in practice by numerically differentiating $\ln L$).
This method is implemented in \texttt{DownhillWLSFitter} (without ECORR) and \texttt{DownhillGLSFitter} (with ECORR), as mentioned in Table \ref{tab:fitters}.

The pulse phases corresponding to the observed TOAs are computed with respect to a fiducial TOA.
Since the choice of this fiducial TOA is arbitrary, there can be an overall phase offset between the measured TOAs and the fiducial TOA.
Traditionally, this offset has been taken care of by subtracting a weighted mean from the timing residuals.
Unfortunately, this procedure is inadequate when correlated noise (including ECORR) is present since the weighted mean does not account for the correlated noise.
Hence, we treat this offset as a free parameter while fitting for correlated noise parameters (this is implemented in the \texttt{PhaseOffset} component, see Table \ref{tab:model-components}).

\updated{While the iterative method described in this section effectively maximizes the likelihood function over both timing and noise parameters, the estimated parameter uncertainties may be underestimated because the timing parameters are kept fixed while estimating noise parameters, and vice versa.
This effect is usually small since most timing parameters are determined by the entire dataset while the white noise and ECORR are time-uncorrelated.
Nevertheless, this caveat can become important while employing timing parameters whose effect is time-localized, such as epoch-wise DM measurements using \texttt{DispersionDMX} \citep[e.g.][]{TarafdarNobleson+2022}.} 

We note here in passing that a white noise characterization utility is available in \tempotwo{} via the \texttt{fixData} plugin \citep{Hobbs2014}.
However, the \pint{} functionality described herein differs from the \texttt{fixData} plugin in the following ways.
(1) The \texttt{fixData} plugin cannot estimate ECORR parameters.
(2) It does not provide uncertainties associated with noise parameter estimates.
(3) Noise parameter estimation in \pint{} is integrated into the usual interactive pulsar timing procedures, whereas \texttt{fixData} is to be run as a separate step.


\subsection{Fitting for time-correlated noise parameters}
\label{sec:rednoisefit}
Recalling the basis given in equation \eqref{eq:fourier-basis}, we can express the deterministic Fourier representation of a time-correlated noise in the following way:
\begin{align}    R(t_{i})&=\sum_{j=1}^{N_{\text{harm}}}\left(\frac{\nu_{\text{ref}}}{\nu_{i}}\right)^{\alpha}\left\{ a_{j}\cos\left[\frac{2\pi j\left(t_{i}-t_{\text{ref}}\right)}{T_{\text{span}}}\right]
+b_{j}\sin\left[\frac{2\pi j\left(t_{i}-t_{\text{ref}}\right)}{T_{\text{span}}}\right]\right\} \,.
\label{eq:fourier-rn-det}
\end{align}
We estimate the coefficients $a_j$ and $b_j$ by treating them as free parameters and fitting them simultaneously with the timing model parameters.
Note that the noise components with frequency less than $T_{\text{span}}^{-1}$ will be absorbed while fitting for the pulsar rotational frequency and its \updated{first derivative} in the case of \updated{the ARN}, and into the \updated{first two DM derivatives} in the case of \updated{the DMN} \updated{when the spectral index is less than 7.
In the case of steeper spectra ($\gamma>7$), higher derivatives of rotational frequency and/or DM may be required \citep{BlandfordNarayanRomani1984, vanHaasterenLevin2012, VanHaasterenVallisneri2014a, KeithNitu2023}.}

The power spectrum of a time-correlated noise component is often modeled using a power law function of the form \citep{LentatiHobson+2014}
\begin{align}
    P(f_{j})=\left\langle a_{j}^{2}\right\rangle =\left\langle b_{j}^{2}\right\rangle=\frac{A^{2}}{12\pi^{2}f_{\text{yr}}^{3}T_\text{span}}\left(\frac{f_{\text{yr}}}{f_{j}}\right)^{\gamma}\,,
\end{align}
where $A$ is the power law amplitude, $\gamma$ is the spectral index (not to be confused with the chromatic index $\alpha$), and $f_\text{yr}=1\,\text{yr}^{-1}$.
In our framework, the power law parameters $A$ and $\gamma$ can be estimated by maximizing the log-likelihood function
\begin{align}
    \ln\Lambda&=-\frac{1}{2}\sum_{j=1}^{N_{\text{harm}}}\left\{ \frac{\hat{a}_{j}^{2}}{P(f_{j})+\epsilon_{aj}^{2}}+\frac{\hat{b}_{j}^{2}}{P(f_{j})+\epsilon_{bj}^{2}}
+\ln\left[P(f_{j})+\epsilon_{aj}^{2}\right]+\ln\left[P(f_{j})+\epsilon_{bj}^{2}\right]\right\} \,,
    \label{eq:spectral-lnl}
\end{align}
where $\hat{a}_{j}$ and $\hat{b}_{j}$ are the maximum-likelihood estimates of the Fourier coefficients obtained by fitting equation \eqref{eq:fourier-rn-det} to the TOAs \updated{as part of the timing model}, and $\epsilon_{aj}$ and $\epsilon_{bj}$ are the measurement uncertainties thereof.
\updated{Here, we are modeling the measured Fourier amplitudes $\hat{a}_{j}$ and $\hat{b}_{j}$ as Gaussian random variables with zero mean whose variance is the sum of the intrinsic variance $P(f_i)$ and the measurement variance $\epsilon_{aj}^{2}$ or $\epsilon_{bj}^{2}$.}
The uncertainties in $A$ and $\gamma$ measurements can be estimated using the Hessian of $\ln\Lambda$ similar to how the EFAC, EQUAD, and ECORR uncertainties are estimated in Subsection \ref{sec:whitenoisefit}.
We note in passing that the likelihood function given in equation \eqref{eq:spectral-lnl} is analogous to the frequency-domain likelihood given in \citet{LaalLamb+2023}.

The likelihood function \eqref{eq:spectral-lnl} will, in general, provide different parameter estimation results than the likelihood function \eqref{eq:gls-like} (including the GP representation of the time-correlated noise), since the latter acts directly on the timing residuals whereas the former acts on the estimated Fourier coefficients.
Additionally, the interpretation of $P(f_j)$ as the variance of the Fourier coefficients $a_j$ and $b_j$ are imposed in equation \eqref{eq:spectral-lnl} \textit{after} estimating $a_j$ and $b_j$, whereas it is imposed as a \textit{prior} distribution in the Bayesian analysis involving the GP representation.
Nevertheless, we expect their results to be broadly consistent.

These computations are implemented in the \texttt{pint.utils.plrednoise\_from\_wavex()} function for the ARN and in the  \texttt{pint.utils.pldmnoise\_from\_dmwavex()} function for the DMN.

\updated{Certain caveats regarding the issues that may arise while applying these techniques to real datasets are in order.
The astrometric parameters of the pulsar, which determine the solar system R{\o}mer delay with a 1-year period, are covariant with the Fourier coefficients of the ARN corresponding to the frequency closest to 1 yr$^{-1}$ (and possibly its higher harmonics). 
Similarly, some of the binary orbital parameters will be covariant with the ARN Fourier coefficients corresponding to the frequency closest to the orbital frequency  (and possibly its higher harmonics), if the orbital period is long enough to be between $T_\text{span}/N_{\text{harm}}$ and $T_\text{span}$.
If the solar wind electron density is included in the timing model as a free parameter, it can be covariant with the Fourier coefficients of the DMN corresponding to the frequency closest to 1 yr$^{-1}$.
We recommend excluding such Fourier coefficients while estimating the spectral parameters using equation \eqref{eq:spectral-lnl} to avoid bias.
Further, our technique assumes that the Fourier basis given in equation \eqref{eq:fourier-basis} is orthogonal.
Although this assumption is exactly true only in the case of uniformly sampled datasets, it will remain approximately true as long as there are no large data gaps or discrepancies in data quality over time.
How to handle cases where this assumption breaks down will be explored in a future work.
Finally, there can be large covariances between \texttt{WaveX} and \texttt{DMWaveX} coefficients of the same frequency when the observing bandwidth is low or when only high-frequency observations are available.}

\updated{A similar utility for estimating the red noise spectrum is available in \tempotwo{} via the \texttt{spectralModel} plugin \citep{ColesHobbs+2011}. 
In terms of usage, it differs from our method in the following ways. 
(1) \texttt{spectralModel} only performs the spectral characterization of the ARN, whereas our method also handles DMN. 
(2) In \texttt{spectralModel}, the optimal spectral index is estimated by manual iteration whereas our method estimates it automatically. 
(3) \texttt{spectralModel} uses a model which is a power law at high frequencies but saturates to a constant at low frequencies, whereas our method uses a simple power law model. 
(4) \texttt{spectralModel} does not provide uncertainties on the estimated spectral parameters.}

\section{Model comparison}
\label{sec:modelcomp}

An important problem in creating and refining a timing model is determining what configuration of model components produces the optimal fit to the data.
In a Bayesian setting, this comparison is accomplished by computing Bayes factors through techniques such as nested sampling \citep{AshtonBernstein+2022} and product space sampling \citep{HeeHandley+2015}.
Since \pint{} follows a frequentist maximum-likelihood approach to parameter estimation, these techniques are not suitable to be used within \pint{}, in addition to being too computationally expensive to be used during interactive pulsar timing.
\updated{Hence, we implemented the AIC in \pint{} to perform model comparison.}
The AIC is defined as
\begin{align}
    \text{AIC} = 2q - 2\ln \hat{L}\,,
\end{align}
where $q$ is the total number of free parameters including the timing model parameters and the noise model parameters, and $\hat{L}$ is the maximum value of the likelihood for the given model.
Given multiple models applied to the same data, the preferred model is the one that minimizes AIC (with a minimum value $\text{AIC}_{\min}$).
The $i$th model can be said to be $\exp[{\text{AIC}_{\min} - \text{AIC}_i}]$ times as probable as the favored model in minimizing information loss
(the quantity $\text{AIC}_{\min} - \text{AIC}_i$ is usually referred to as the AIC difference).
See, e.g., \citet{BurnhamAnderson2004} for a detailed description and interpretation of AIC. 
AIC can be computed in \pint{} using the \texttt{akaike\_information\_criterion()} function available in the \texttt{pint.utils} module. 

The application of AIC in selecting the applicable noise components and the estimation of noise parameters is demonstrated in the next section.
Although these examples only involve noise components, AIC can be used for other types of comparisons, e.g., between different binary models.


\section{Simulation studies}
\label{sec:sims}

In this Section, we \updated{demonstrate} the efficacy of our maximum-likelihood noise estimation methods by applying them to simulated datasets and comparing the results with Bayesian estimates obtained using the \texttt{ENTERPRISE} package (see Table \ref{tab:priors} for the prior distributions used in the Bayesian analyses).
In what follows, we investigate \updated{five} cases: (a) white noise-only, (b) white noise + ECORR, (c) ARN-only, (d) DMN-only, \updated{and (e) ARN + DMN}.

The simulations are performed using the \texttt{pint.simulation} module of \pint{} (see Subsection \ref{sec:pintsim}). 
Each simulation corresponds to a fictitious isolated pulsar with a spin frequency of 100 Hz and spin frequency derivative of $-10^{-15}$ Hz$^2$, located at right ascension 05$^h$00$^m$00$^s$ and declination 15$^\circ$00$'$00$''$, with a dispersion measure of 15 pc/cm$^3$.
The solar system delays are estimated using the DE440 solar system ephemeris \citep{ParkFolkner+2021}.
Each simulation contains 2000 narrowband TOAs taken at 250 equally spaced epochs spanning an MJD range of 53000--57000 taken at the Green Bank Telescope (GBT).
Each epoch contains TOAs from eight equally spaced frequency subbands in the 500--1500 MHz range.
The unscaled TOA uncertainties ($\sigma_i$) are drawn uniformly from the interval 0.5--2.0 $\mu$s. 
Table \ref{tab:sims} lists the noise parameters injected into each simulation.

\updated{These simulations are intentionally kept simplistic so that the demonstrations focus on one aspect at a time and remain clear. 
Please see Section \ref{sec:b1855+09} for an application of our method on a real dataset.}

\begin{table*}[]
\centering
\begin{tabular}{c|c|c|c}
\hline
\textbf{Simulation \#}   & \textbf{Noise Parameter} & \textbf{Units} & \textbf{Injected Value} \\ \hline
\multirow{2}{*}{(a)} & EFAC[tel gbt]             &                & 1.3                     \\ 
                      & EQUAD[tel gbt]            & $\mu$s             & 0.9                     \\ \hline
\multirow{2}{*}{(b)} & EFAC[tel gbt]             &                & 1.3                     \\ 
                      & ECORR[tel gbt]            & $\mu$s             & 0.9                     \\ \hline
\multirow{3}{*}{(c)} & TNREDAMP                 &                &       $-$13                  \\ 
                      & TNREDGAM                 &                &   3.5                      \\ 
                      & TNREDC                   &                &   30                      \\ \hline
\multirow{3}{*}{(d)} & TNDMAMP                  &                &   $-$13.5                      \\ 
                      & TNDMGAM                  &                &   4.0                      \\ 
                      & TNDMC                    &                &   30                      \\ \hline
                \multirow{6}{*}{(e)} & TNREDAMP                 &                &       $-$13                  \\ 
                      & TNREDGAM                 &                &   3.5                      \\ 
                      & TNREDC                   &                &   30                      \\ 
                      & TNDMAMP                  &                &   $-$13.5                      \\ 
                      & TNDMGAM                  &                &   4.0                      \\ 
                      & TNDMC                    &                &   30                      \\ 
                      \hline
\end{tabular}
\caption{Noise parameters injected into various simulations. 
\updated{The parameter names used here correspond to how they are represented in `par' files.}
TNREDAMP, TNREDGAM, and TNREDC represent the GP log amplitude, spectral index, and number of harmonics \updated{of the ARN}. 
TNDMAMP, TNDMGAM, and TNDMC represent the GP log amplitude, spectral index, and number of harmonics \updated{of the DMN}.}
\label{tab:sims}
\end{table*}

\begin{table}
    \centering
    \begin{tabular}{c|c|c}
        \hline
        \textbf{Noise Parameter} & \textbf{Units} & \textbf{Prior Distribution} \\\hline
        EFAC & & Uniform[0.5, 2.0] \\
        EQUAD & $\mu$s & Uniform[0.01, 100] \\
        ECORR & $\mu$s & Uniform[0.01, 100] \\
        TNREDAMP &  & Uniform[$-$20, $-$11] \\
        TNREDGAM &  & Uniform[0, 7] \\
        TNDMAMP &  & Uniform[$-$20, $-$11] \\
        TNDMGAM &  & Uniform[0, 7] \\
        \hline
    \end{tabular}
    \caption{Prior distributions used for the Bayesian noise analysis using \enterprise{} in Sections \ref{sec:sims} and \ref{sec:b1855+09}.
    Note that EFAC, EQUAD, and ECORR are expressed in linear scale whereas TNREDAMP and TNDMAMP are expressed in log scale in this table.
    The parameter names used here correspond to how they are represented in `par' files.}
    \label{tab:priors}
\end{table}

\subsection{\updated{Simulation (a): White noise-only, EFAC \& EQUAD}}
\label{sec:sim1}
In this simulation, we modify the TOA uncertainties by an EFAC and an EQUAD \updated{(see Table \ref{tab:sims})}.
We fit the generated TOAs for both timing and noise parameters starting from an initial model with EFAC=1 and EQUAD=$10^{-3}$ $\mu$s (a small positive value close to 0). 
To determine which noise parameters are necessary, we explore four versions of this fit, where (i) the parameters are frozen at EFAC=1 and EQUAD=0, (ii) EFAC is allowed to vary but EQUAD is frozen at 0,  (iii) EQUAD is allowed to vary but EFAC is frozen at 1, and (iv) both EFAC and EQUAD are allowed to vary.
The AIC differences for these configurations are listed in Table \ref{tab:sim1-aic}.

\begin{table}
\centering
\begin{tabular}{l|c}
    \hline
   Configuration  & AIC Difference \\ \hline
   Free EFAC, Free EQUAD  & 0 \\
   Free EFAC, EQUAD=0 & 44 \\
   EFAC=1, Free EQUAD & 198 \\
   EFAC=1, EQUAD=0 & 5191\\
   \hline
\end{tabular}
\caption{AIC differences for different noise model configurations for \updated{simulation (a) with EFAC \& EQUAD (white noise-only).} See Subsection \ref{sec:sim1} for more details.}
\label{tab:sim1-aic}
\end{table}

\begin{figure}[t!]
    \centering
    \begin{tabular}{cc}
    \includegraphics[width=0.5\textwidth,trim={0 -3cm 0 0},clip]{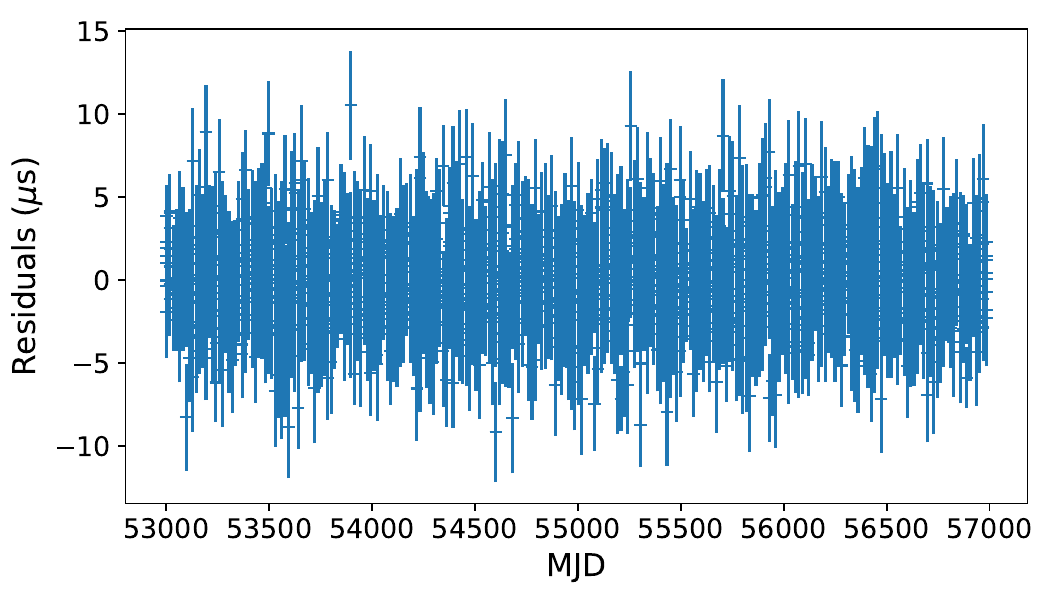}
    &\includegraphics[width=0.5\textwidth,trim={0 0 0 0},clip]{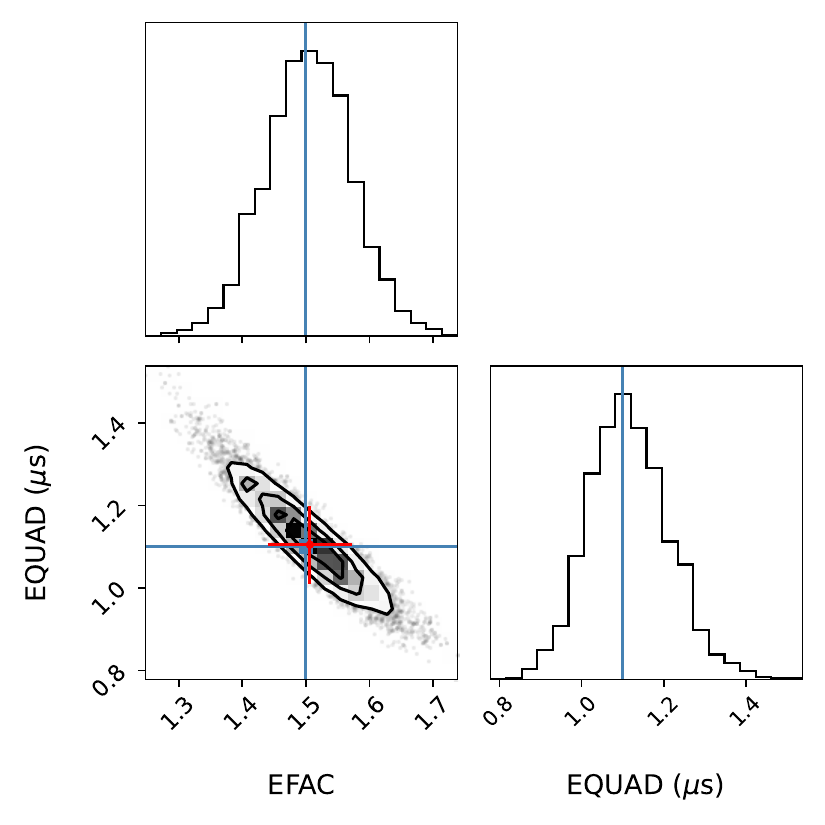}
    \\(a) & (b)\\
    \multicolumn{2}{c}{\includegraphics[width=0.5\textwidth]{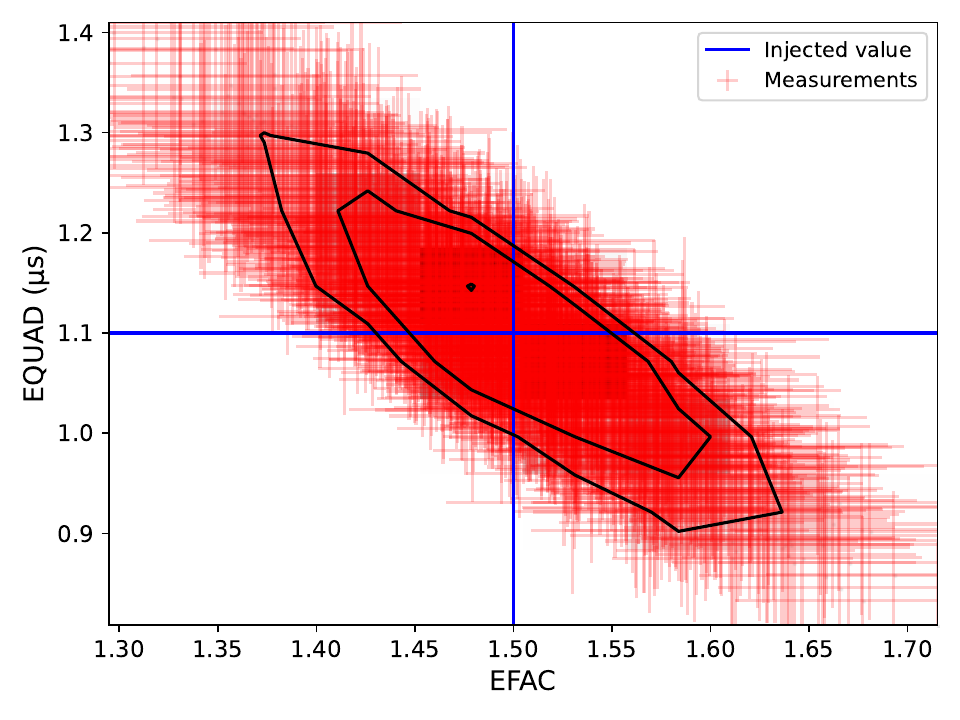}} \\
    \multicolumn{2}{c}{(c)}
    \end{tabular}
    \caption{Timing residuals (panel a) and parameter estimation results (panel b) for the \updated{simulation (a) with EFAC + EQUAD} (see Subsection \ref{sec:sim1}). 
    In panel (b), the blue lines show the injected values, the red point with error bars shows the maximum likelihood estimate obtained using \pint{}, and the corner plot (black) shows the posterior distribution obtained from the \enterprise{} analysis.
    \updated{Panel (c) shows the result of 500 repetitions of this simulation; red points with error bars show the maximum-likelihood estimates for each realization, the blue lines indicate the injected values and the black contours indicate the distribution of the maximum-likelihood values.}}
    \label{fig:sim1}
\end{figure}

Clearly, both the EFAC and the EQUAD are required to model the white noise in this dataset.
The comparison of the measured EFAC and EQUAD values with the injected values is shown in Figure \ref{fig:sim1}(b).
Figure \ref{fig:sim1}(b) also shows the posterior distribution obtained from a Bayesian analysis performed using \enterprise{} and \texttt{PTMCMCSampler} \citep{JohnsonMeyers+2024}; see Table \ref{tab:priors} for the prior distributions used. 
We see that the maximum likelihood EFAC and EQUAD estimates are consistent with the injected values and the Bayesian estimates within error bars.

\updated{To ensure that the results are robust, we generated 1000 instances of this simulation and estimated the maximum-likelihood noise parameters for each instance. 
These measurements, along with the injected values, are shown in Figure \ref{fig:sim1}(c).
Comparing the measured EFAC \& EQUAD values with their injected values using Student's t-test, we find no evidence that the estimates are biased (the $p$-values are around 0.8 for EFAC and 0.9 for EQUAD).
The covariance between EFAC and EQUAD seen in the Bayesian posterior distribution in Figure \ref{fig:sim1}(b) is also seen in Figure \ref{fig:sim1}(c).}

~

\subsection{\updated{Simulation (b): EFAC \& ECORR}}
\label{sec:sim2}

In this simulation, we modify the TOA uncertainties by an EFAC and include an ECORR noise component. 
Similar to the white noise-only case, we fit the generated TOAs for both timing and noise parameters starting from an initial model with EFAC=1 and ECORR=$10^{-3}$. 
We explore four versions of this fit, where each noise parameter is allowed to be free or frozen.
The AIC differences for these configurations are listed in Table \ref{tab:sim2-aic}.

\begin{table}
\centering
\begin{tabular}{l|c}
    \hline
   Configuration  & AIC Difference \\ \hline
   Free EFAC, Free ECORR  & 0 \\
   Free EFAC, ECORR=0 & 228 \\
   EFAC=1, Free ECORR & 331 \\
   EFAC=1, ECORR=0 & 1238\\
   \hline
\end{tabular}
\caption{AIC differences for different noise model configurations for \updated{simulation (b) with EFAC \& ECORR}. See Subsection \ref{sec:sim2} for more details.}
\label{tab:sim2-aic}
\end{table}

\begin{figure}[ht!]
    \centering
    \begin{tabular}{cc}
    \includegraphics[width=0.5\textwidth,trim={0 -3cm 0 0},clip]{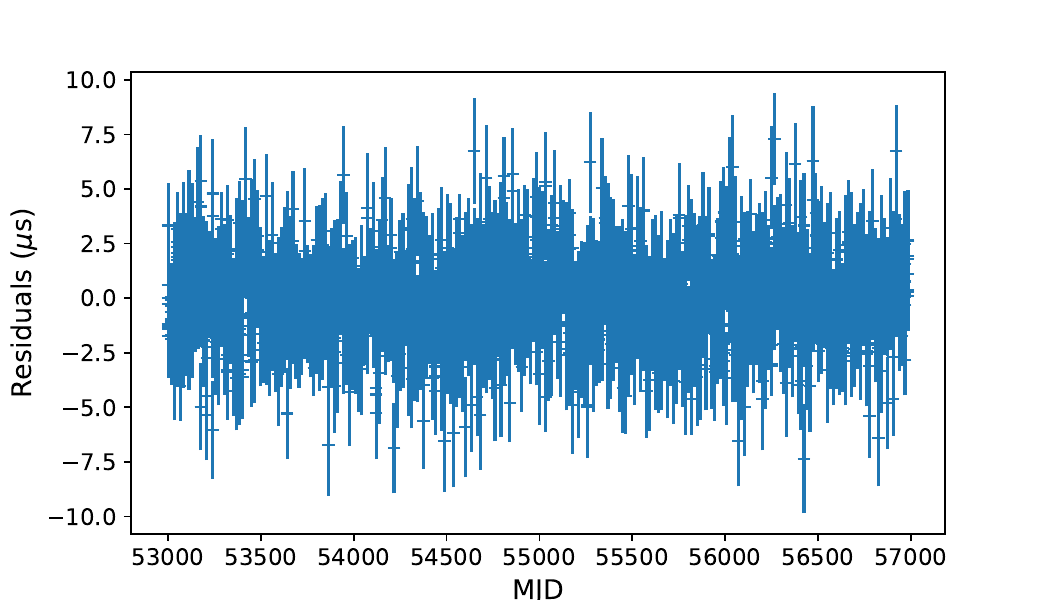}
         & \includegraphics[width=0.5\textwidth,trim={0 0cm 0 0cm},clip]{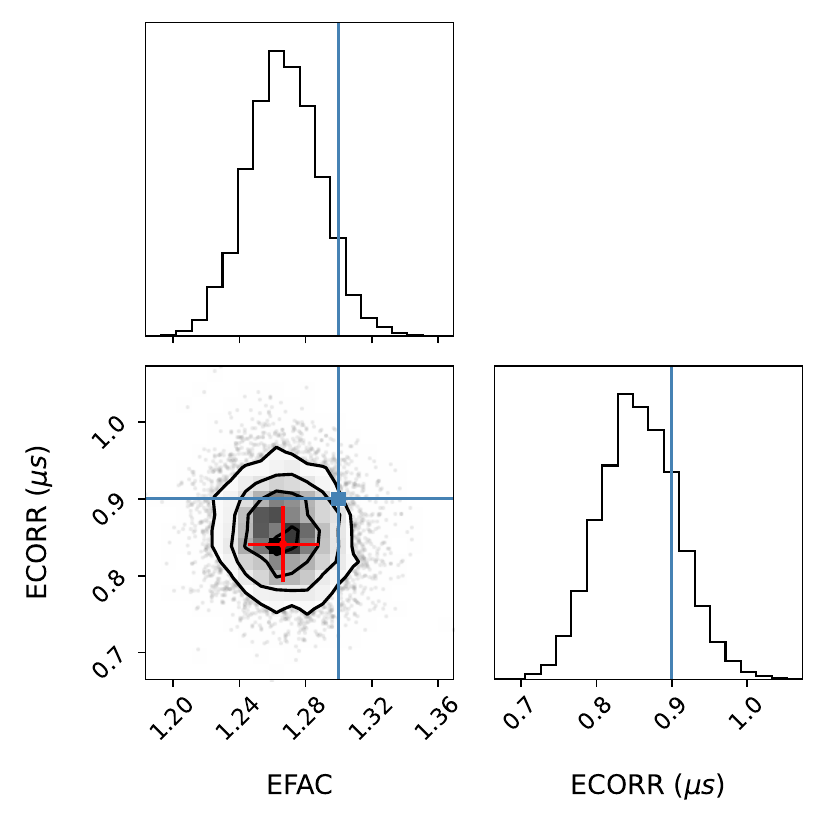}\\
        (a)&(b)  \\
    \multicolumn{2}{c}{\includegraphics[width=0.5\textwidth]{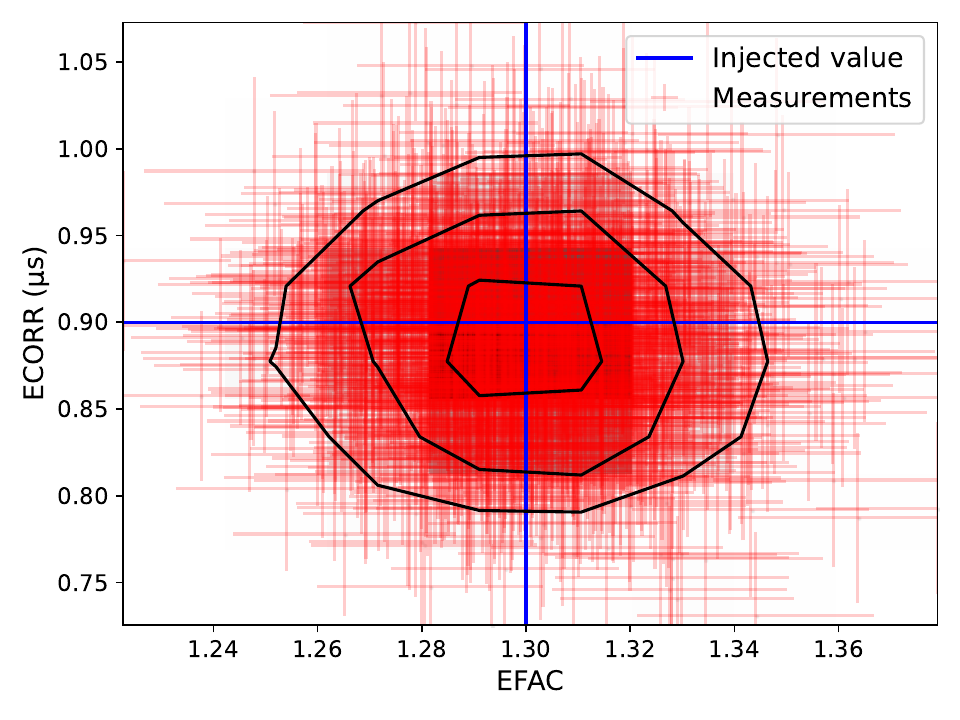}}\\
    \multicolumn{2}{c}{(c)}
        \end{tabular}
    \caption{The timing residuals (panel a) and the parameter estimation results (panel b) for \updated{simulation (b) with EFAC \& ECORR} (see Subsection \ref{sec:sim2}). 
    \updated{Panel (c) shows the results of 500 repetitions of this simulation.}
    The plotting conventions are identical to Figure \ref{fig:sim1}.}
    \label{fig:sim2}
\end{figure}

Clearly, both the EFAC and the ECORR are required to model the noise in this dataset.
The comparison of the measured EFAC and ECORR values with the injected values is shown in Figure \ref{fig:sim2}(b), along with the posterior distribution obtained from a Bayesian \texttt{ENTERPRISE} analysis; 
see Table \ref{tab:priors} for the prior distributions used.
We see that the maximum likelihood estimates are consistent with the Bayesian estimates within error bars whereas both estimates are offset from the injected values.

\updated{To check the robustness of our method we ran 1000 instances of this simulation and estimated the parameters for each.
The results of this exercise are shown in Figure \ref{fig:sim2}(c). 
Comparing the measured values of the EFAC \& ECORR with their injected values using Student's t-test, we find that there is no evidence for bias in the EFAC estimations, whereas ECORRs are underestimated by around $0.2\sigma$ (the $p$-values are around $0.8$ and $3\times 10^{-11}$ respectively).
The parameter estimates in Figure \ref{fig:sim2}(b), while more than 1$\sigma$ away from the injected value, are still within the distribution seen in Figure \ref{fig:sim2}(c), implying that the apparent offset can be attributed to random chance.
}

\subsection{\updated{Simulation (c): ARN-only}}
\label{sec:sim3}

\begin{figure*}[ht!]
    \centering
    \begin{tabular}{cc}
       \includegraphics[width=0.5\textwidth,trim={0 0 0 0},clip]{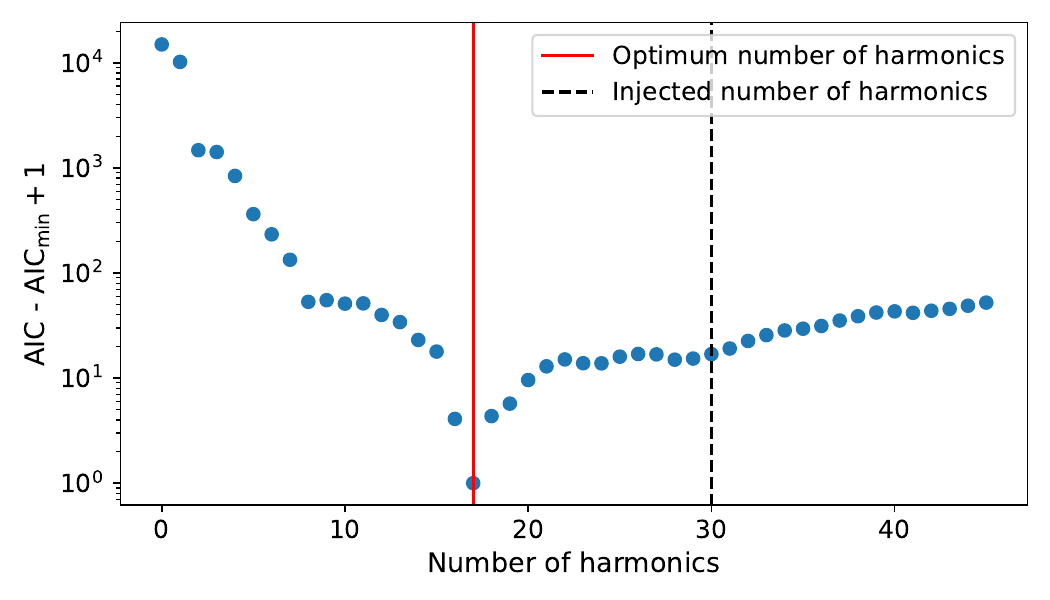}  & 
       \includegraphics[width=0.5\textwidth,trim={0 0cm 0 0cm},clip]{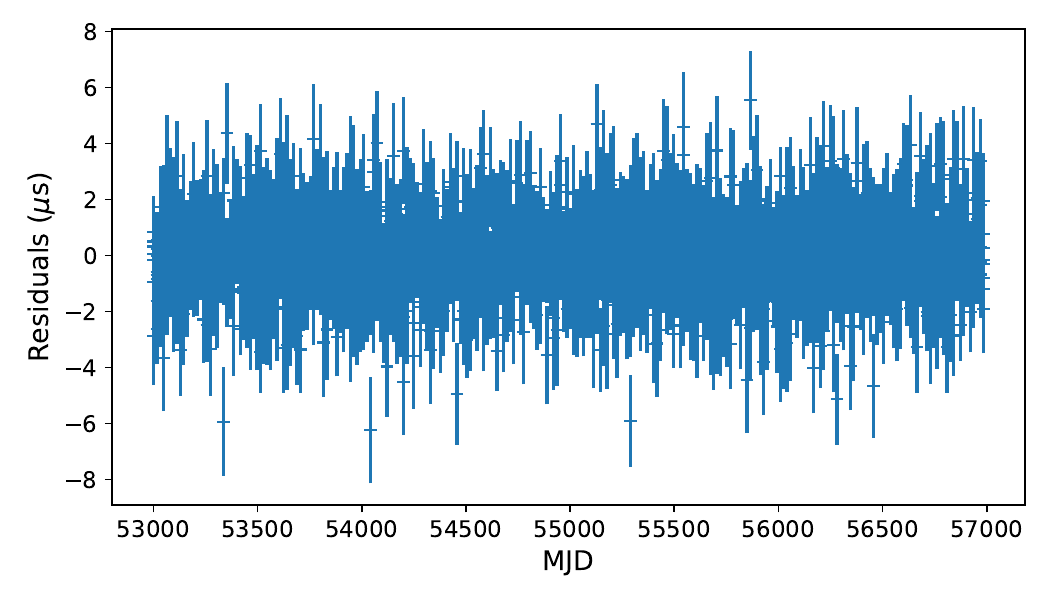}\\
       (a) & (b)\\
       \includegraphics[width=0.5\textwidth, trim={0 -1cm 0 0},clip]{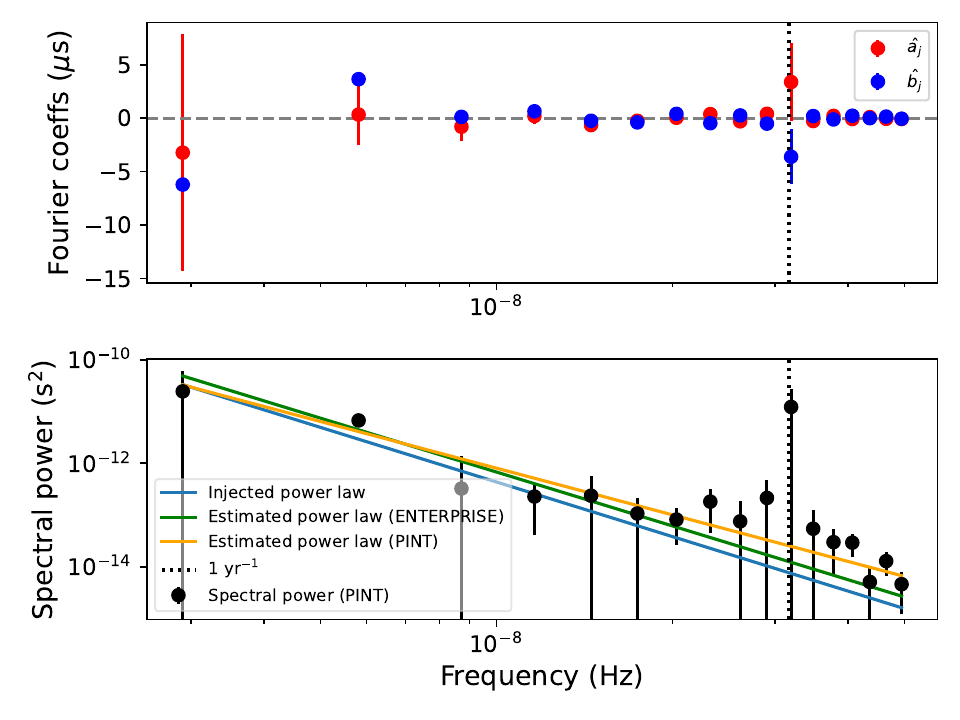}
         & 

        \includegraphics[width=0.5\textwidth,trim={0 0cm 0 0cm},clip]{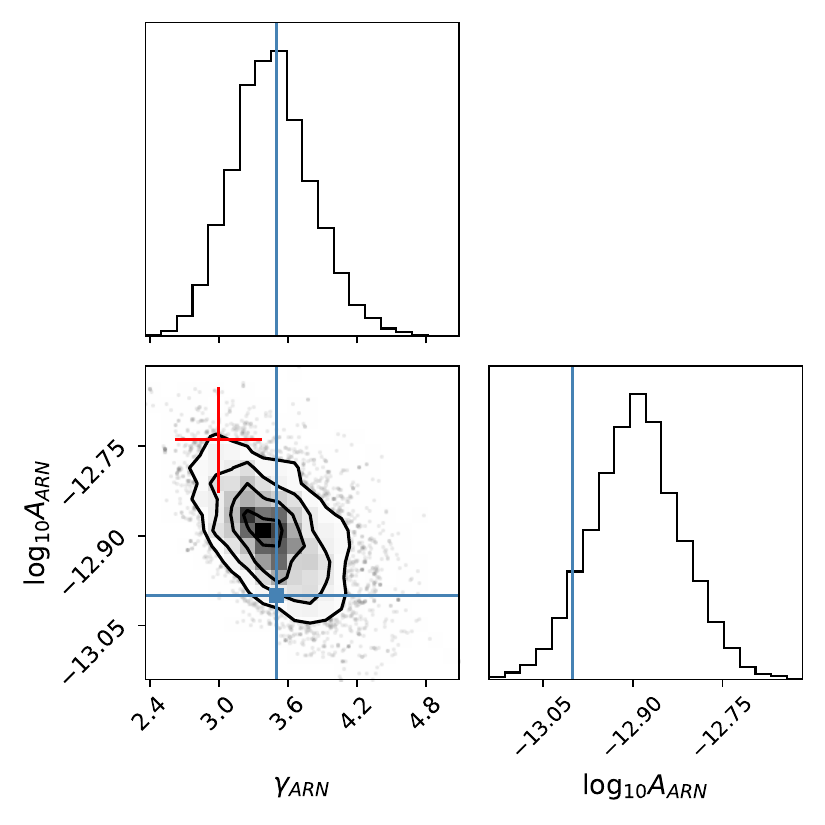}\\
        (c) & (d) \\        \multicolumn{2}{c}{\includegraphics[width=0.5\textwidth]{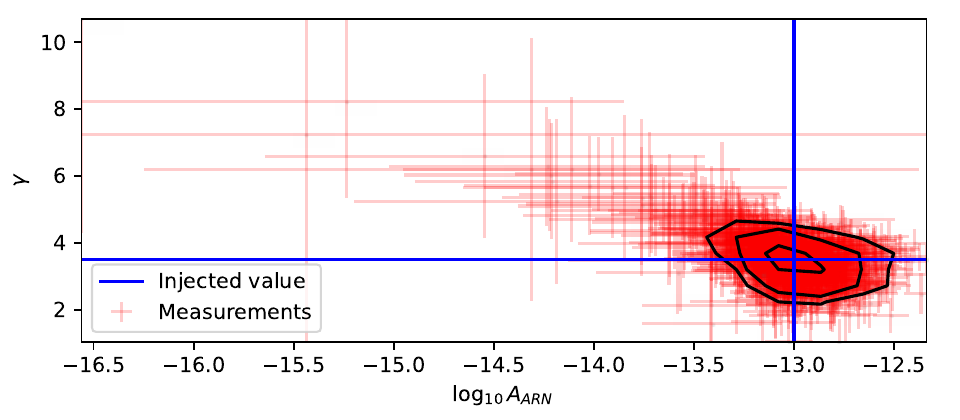}} \\
        \multicolumn{2}{c}{(e)}  
    \end{tabular}
    \caption{\textit{Panel (a)} shows the $\text{AIC}-\text{AIC}_{\min} + 1$ values for the different number of harmonics for \updated{simulation (c) with only the ARN} (see Subsection \ref{sec:sim3}). 
    A red vertical line indicates the optimum number of harmonics. 
    The injected value is indicated by a black dashed vertical line. 
    The Y axis is plotted in a log scale for better visibility (the +1 term is added to $\text{AIC}-\text{AIC}_{\min}$ to enable this). 
    \textit{Panel (b)} shows the post-fit timing residuals. 
    \textit{Panel (c)} shows the parameter estimation results. 
    The top sub-panel shows the maximum likelihood estimates of the Fourier sine (blue points) and cosine (red points) coefficients $\hat{a}_j$ and $\hat{b}_j$ appearing in equation \eqref{eq:fourier-rn-det}.
    The bottom panel shows the corresponding spectral powers (black points) and the power law fit thereof (orange line) obtained by maximizing the likelihood function given in equation \eqref{eq:spectral-lnl}.
    For comparison, the bottom panel also shows the injected power law spectrum (blue line) and the power law spectrum estimated using \texttt{ENTERPRISE} in a Bayesian way (green line). 
    The black dotted line corresponds to 1 yr$^{-1}$. 
    \textit{Panel (d)} shows the comparison of the power law spectral parameters obtained using \pint{} and \enterprise. The plotting conventions here are identical to Figure \ref{fig:sim1}.
    \updated{\textit{Panel (e)} shows the results of 1000 repetitions of this simulation, with identical plotting conventions as Figure \ref{fig:sim1}(c).}}
    \label{fig:sim3}
\end{figure*}

In this simulation, we inject a realization of the ARN component with a power law spectrum, modeled as a Fourier GP (implemented in \pint{} as the \texttt{PLRedNoise} model), into the TOAs.
We model this noise using the \texttt{WaveX} model, where the component frequencies are taken to be the harmonics of a fundamental frequency $T_\text{span}^{-1}$, where $T_\text{span}$ is the total observation span.
To determine the optimal number of harmonics needed to model the noise, we fit the simulated TOAs using different numbers of harmonics and compute the AIC value corresponding to each case.
These AIC values are plotted in Figure \ref{fig:sim3}(a), and the optimum number of harmonics turns out to be 17 (the injected value is 30, see Table \ref{tab:sims}).
The maximum-likelihood estimates for the Fourier coefficients $\hat{a}_i$ and $\hat{b}_i$ are plotted in the top panel of Figure \ref{fig:sim3}(c), along with its power spectrum in the bottom panel.
In Figure \ref{fig:sim3}(c), we see an outlier in the frequency bin containing 1 yr$^{-1}$.
This can be attributed to the covariance of these Fourier coefficients with the \updated{astrometric parameters as discussed in Subsection \ref{sec:rednoisefit}}.
We fit a power law to the estimated Fourier coefficients as described in Subsection \ref{sec:rednoisefit}, estimating the power law amplitude and the spectral index \updated{while ignoring the 1 yr$^{-1}$ bin}.
The best-fit power law model is also shown in the bottom panel of Figure \ref{fig:sim3}(c), along with its Bayesian counterpart obtained using \texttt{ENTERPRISE} as well as the injected power-law spectrum for comparison. 
Further comparison of these spectral parameter estimates are shown in Figure \ref{fig:sim3}(d), where the posterior distribution obtained from the \texttt{ENTERPRISE} analysis and the injected values are also plotted.


\updated{To assess the robustness of our method, we created 1000 instances of this simulation and estimated the parameters for each. 
The distribution of the measured values is plotted in Figure \ref{fig:sim3}(e).
We see that the mean of this distribution is consistent with the injected value (the Student t-test $p$-values are around 0.6 and 0.7 for the log amplitude and the spectral index respectively).
However, the distribution has a long tail towards lower amplitudes and steeper spectral indices.
Based on Figure \ref{fig:sim3}(e), we see that the offset between the injected and measured values seen in Figure \ref{fig:sim3}(d) is consistent with it being due to random chance.}

\newpage

\subsection{\updated{Simulation (d): DMN-only}}
\label{sec:sim4}

\begin{figure*}[ht!]
    \centering
    \begin{tabular}{cc}
       \includegraphics[width=0.5\textwidth]{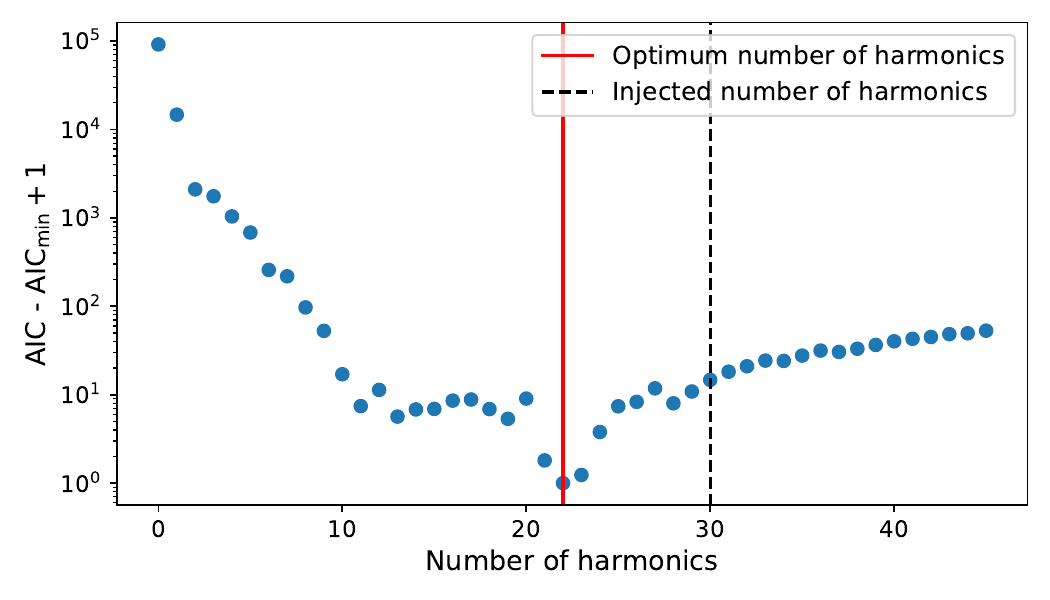}  & 
       \includegraphics[width=0.5\textwidth]{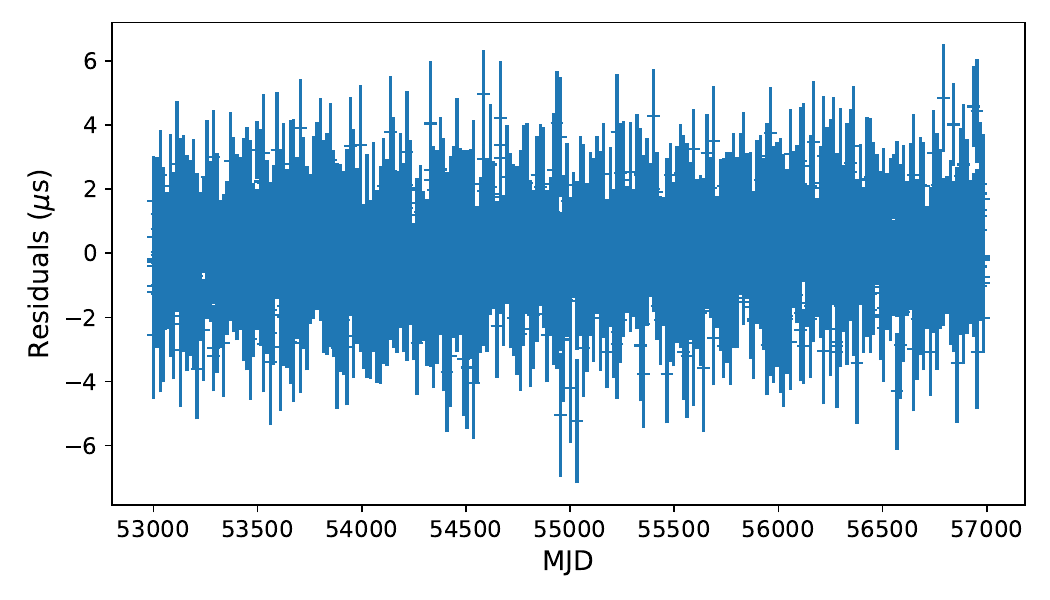}\\
       (a) & (b)\\
       \includegraphics[width=0.5\textwidth, trim={0 -2cm 0 0},clip]{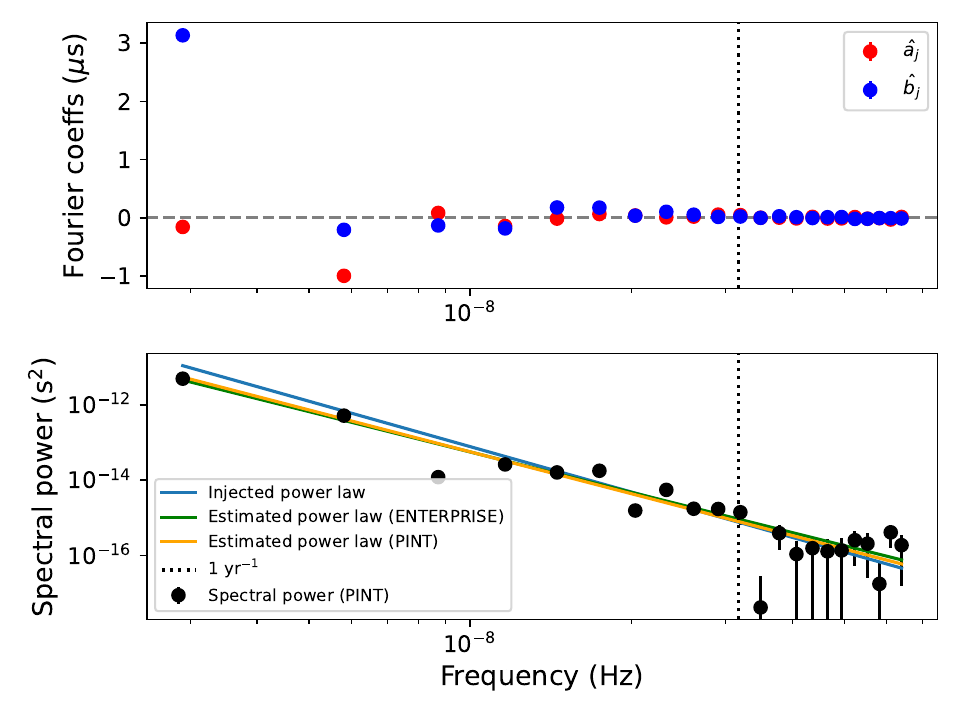}
         & 
        \includegraphics[width=0.5\textwidth,trim={0 0cm 0 0},clip]{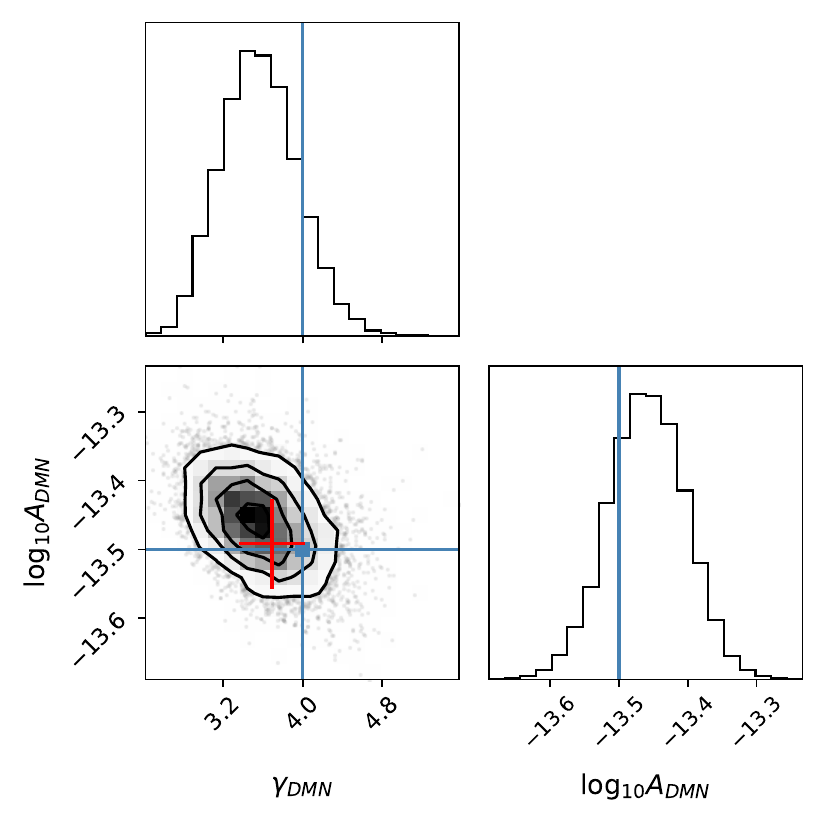}\\
        (c) & (d)\\
        \multicolumn{2}{c}{\includegraphics[width=0.5\textwidth]{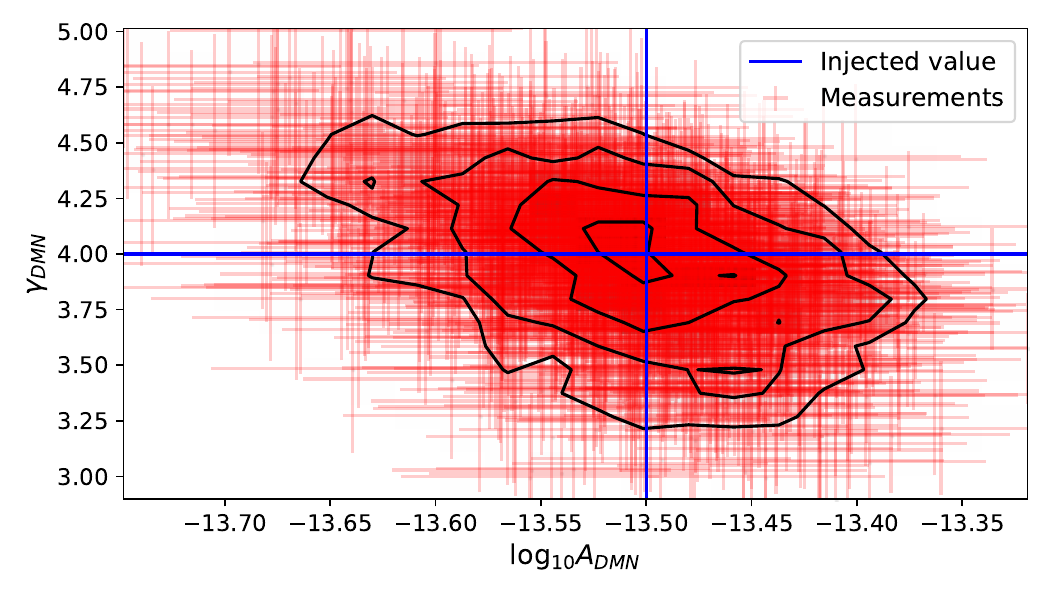}}\\
        \multicolumn{2}{c}{(e)}
    \end{tabular}
    \caption{\textit{Panel (a)} shows the $\text{AIC}-\text{AIC}_{\min} + 1$ values for the different number of harmonics for the \updated{simulation (d) with only the DMN} (see Subsection \ref{sec:sim4}). 
    The plotting conventions are identical to Figure \ref{fig:sim3}(a).  
    \textit{Panel (b)} shows the post-fit timing residuals.  
    \textit{Panel (c)} shows the parameter estimation results. 
    The plotting conventions are identical to Figure \ref{fig:sim3}(c).
    The error bars in the top sub-panel are too small to be visible. 
    \textit{Panel (d)} shows the comparison of the power law spectral parameters obtained using \pint{} and \enterprise. The plotting conventions here are identical to Figure \ref{fig:sim1}.
    \updated{\textit{Panel (e)} shows the results of 1000 repetitions of this simulation, with identical plotting conventions as Figure \ref{fig:sim1}(c).}}
    \label{fig:sim4}
\end{figure*}

In this simulation, we inject a realization of the DMN with a power law spectrum, modeled as a Fourier GP (implemented in \pint{} as the \texttt{PLDMNoise} model), into the TOAs.
We model this noise using the \texttt{DMWaveX} model, where the component frequencies are taken to be harmonics of $T_\text{span}^{-1}$.
We begin by determining the optimal number of harmonics using AIC, and the AIC differences for different numbers of harmonics are plotted in Figure \ref{fig:sim4}(a).
The optimal number of harmonics turns out to be 22 (the injected value is 30).
Figure \ref{fig:sim4}(c) shows the estimated Fourier coefficients in the top panel and the estimated spectral powers, the injected power-law spectrum, and the estimated power-law spectra (both using \pint{} and \texttt{ENTERPRISE}) in the bottom panel.
Figure \ref{fig:sim4}(d) shows the corresponding power law amplitude and spectral index estimates.
We see that the power spectrum estimated using \pint{} and \texttt{enterprise} are consistent with each other as well as with the injected values within the uncertainty levels.

\updated{We repeated this simulation 1000 times and measured the noise parameters in each case, and the results of this exercise are plotted in Figure \ref{fig:sim4}(e).
The means of the measured log-amplitude and spectral index are both offset from the injected values by around $0.1\sigma$ (Student's t-test $p$-values are approximately $4\times 10^{-6}$ and $2\times 10^{-4}$ respectively).}

\subsection{\updated{Simulation(e): ARN \& DMN}}

\begin{figure*}
    \centering
    \begin{tabular}{cc}
       \includegraphics[width=0.5\textwidth,trim={0 0.2cm 0 1.4cm},clip]{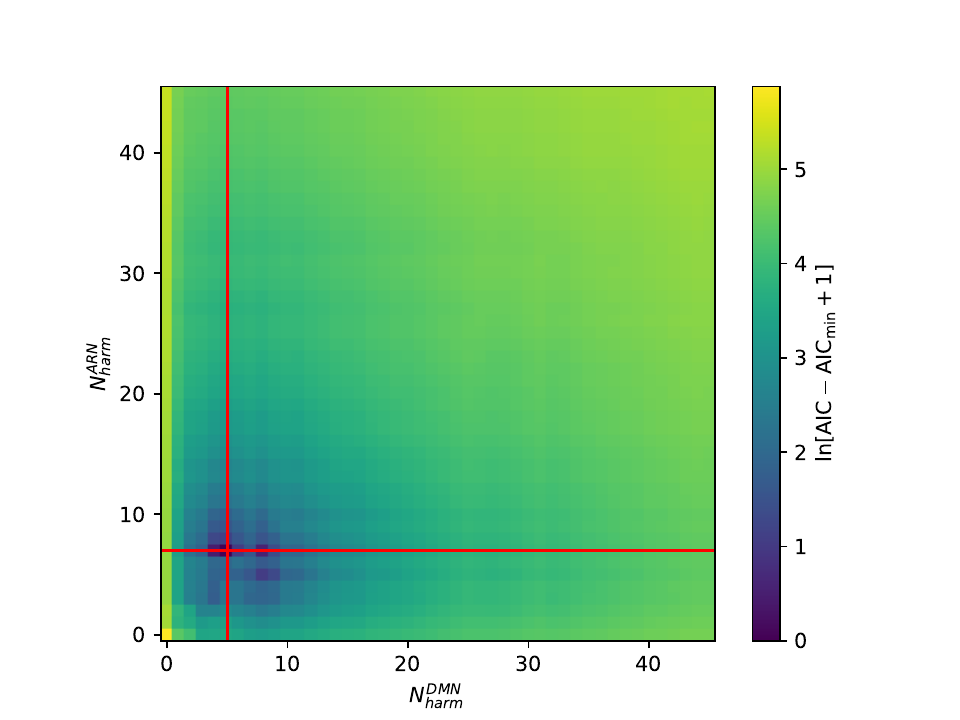}  & \includegraphics[width=0.5\textwidth]{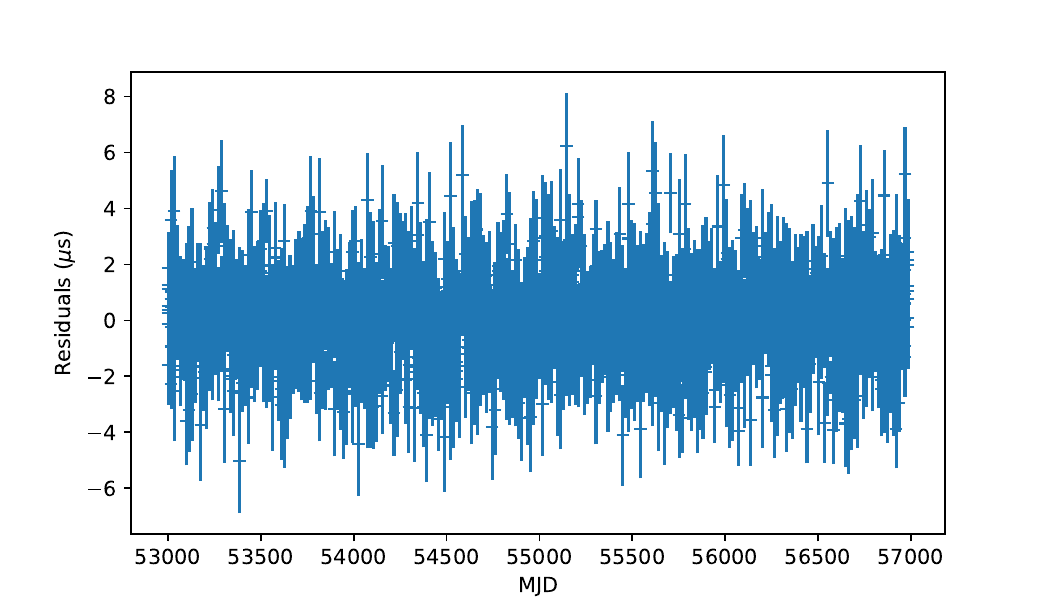}\\
        (a) & (b)\\
        \multicolumn{2}{c}{\includegraphics[width=1\textwidth,trim={0 0.6cm 0 0},clip]{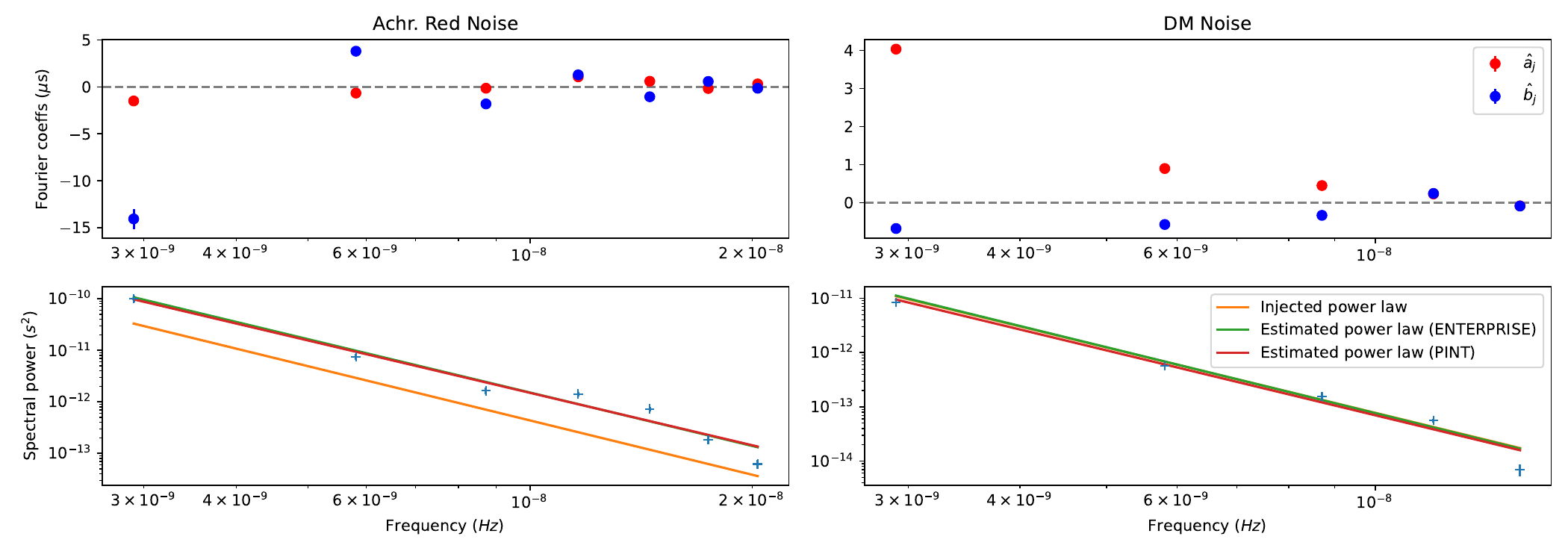}} \\
        \multicolumn{2}{c}{(c)}\\
        \includegraphics[width=0.5\textwidth,trim={0 0.55cm 0 0},clip]{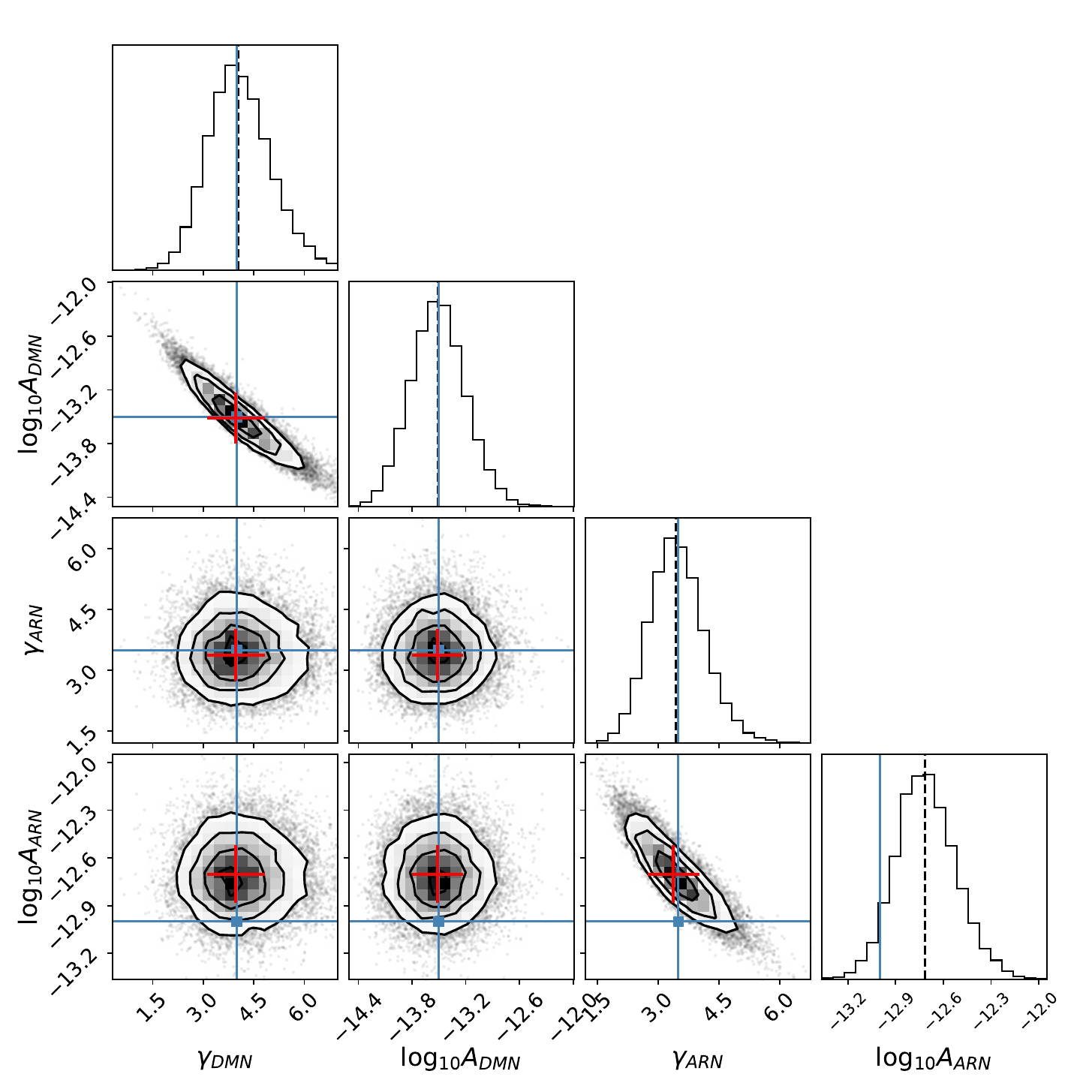} & 
        \includegraphics[width=0.5\textwidth]{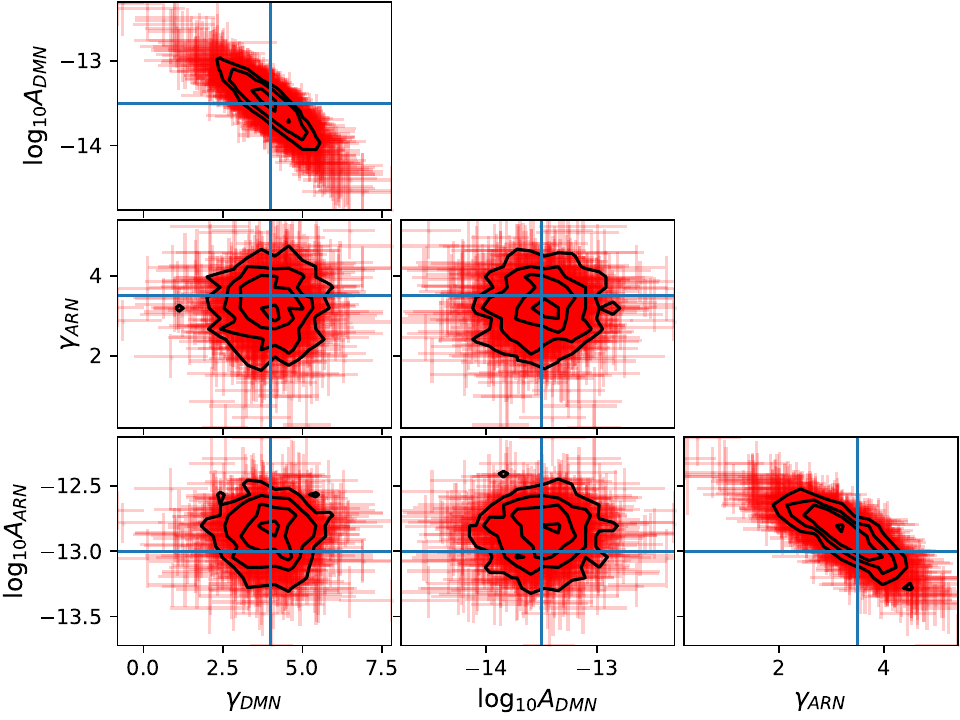}
        \\
       (d)&(e)
    \end{tabular}
    \caption{\updated{Panel (a) shows the AIC differences in log scale as a function of the number of harmonics or ARN and DMN.
    Panel (b) shows the post-fit timing residuals. Panel(c) shows the comparison between the injected and measured power-law spectra. 
    Panel (d) shows the comparison between injected and measured noise parameters.
    Panel (e) shows the results of 1000 repetitions of this simulation.
    The plotting conventions are identical to Figure \ref{fig:sim3}.}}
    \label{fig:sim5}
\end{figure*}

\updated{In this simulation, we inject realizations of the DMN and the ARN with power-law spectra, modeled as Fourier GPs, into the TOAs.
We model this noise using the \texttt{DMWaveX} and \texttt{WaveX} models, where the component frequencies are taken to be harmonics of $T_\text{span}^{-1}$.
We begin by determining the optimal number of harmonics using AIC, and the AIC differences for different numbers of harmonics of the DMN and the ARN are plotted in Figure \ref{fig:sim5}(a) using a color map.
The optimal numbers of harmonics turn out to be 5 for the DMN and 7 for the ARN (the injected values are 30 and 30 respectively).
Figure \ref{fig:sim4}(c) shows the estimated Fourier coefficients in the top panels and the estimated spectral powers, the injected power-law spectra, and the estimated power-law spectra (both using \pint{} and \texttt{ENTERPRISE}) in the bottom panels, for both the DMN (right panels) and the ARN (left panels).
Figure \ref{fig:sim4}(d) shows the corresponding estimates for the power law amplitudes and the spectral indices.
We see that the power spectrum estimated using \pint{} and \texttt{enterprise} are consistent with each other for both DMN and ARN.
However, the ARN spectral parameter estimates are offset from their injected values although the DMN parameter estimates show good agreement with the injected values.}

\updated{We repeated this simulation 1000 times and measured DMN and ARN spectral parameters for each instance.
The results of this exercise are plotted in Figure \ref{fig:sim5}(e).
We find that the mode of the measured DMN parameter values is close to the injected values (spectral index measurements are offset by approximately $0.1\sigma$ from the injected value whereas there is no clear evidence that the log-amplitude measurements are offset; the Student t-test $p$-values are approximately $4\times 10^{-3}$ and $0.2$ respectively).
However, the measured ARN parameter values are more visibly offset from the injected values, namely the spectral index is offset by $0.4\sigma$ ($p\sim 10^{-36}$) and the log-amplitude is offset by $0.7\sigma$ ($p\sim 10^{-83}$).
Further, we find covariance between the measured amplitude and spectral index for ARN and DMN in Figure \ref{fig:sim5}(e) similar to what is seen in the posterior distribution in Figure \ref{fig:sim5}(e).}

\updated{These results indicate that the correlated noise spectral parameter estimates can be biased when both ARN and DMN are present in a dataset.}

\section{Application to PSR B1855+09}
\label{sec:b1855+09}

We now proceed to demonstrate and test our methods on the NANOGrav 9-year (NG9) narrowband dataset of PSR B1855+09.
This dataset contains 4005 TOAs taken using the Arecibo telescope during 2004--2013 \citep{ArzoumanianBrazier+2015}.
It was chosen because it is distributed as an example dataset together with the \pint{} source code.

We used the `par' file from the NG9 dataset as a starting point for our analysis after removing the DMX parameters representing piecewise constant DM variations.
We begin our analysis by fitting the timing model parameters, including
\begin{enumerate}[label=(\alph*)]
    \item Astrometric parameters (sky location, proper motion, and parallax),
    \item Dispersion measure and its time derivatives (up to the second derivative),
    \item Pulsar binary parameters for the \texttt{BinaryDD} model (binary period and period derivative, projected semi-major axis of the pulsar orbit, eccentricity, argument of periapsis, epoch of periapsis passage, companion mass, and orbital sine-inclination),
    \item Frequency-dependent profile evolution parameters (up to third order),
    \item Pulsar rotational frequency and its derivative,
    \item Phase jump between different receivers (\textsl{430} and \textsl{L-wide}),
    \item Overall phase offset.
\end{enumerate}

Thereafter, we included the \texttt{DMWaveX} \updated{and \texttt{WaveX}} models with \updated{40 harmonics each} to model the \updated{DMN and ARN} along with EFAC, EQUAD, and ECORR parameters for each observing system (receiver-backend combinations, denoted as \textsl{430\_ASP}, \textsl{430\_PUPPI}, \textsl{L-wide\_ASP}, and \textsl{L-wide\_PUPPI}).
We have used $T_\text{span}^{-1}$ as the fundamental frequency of the \texttt{DMWaveX} \updated{and \texttt{WaveX}} components.

\begin{figure*}
\centering
\noindent\makebox[\textwidth]{
\begin{tabular}{c}
\begin{tabular}{ccc}
\includegraphics[width=0.33\textwidth]{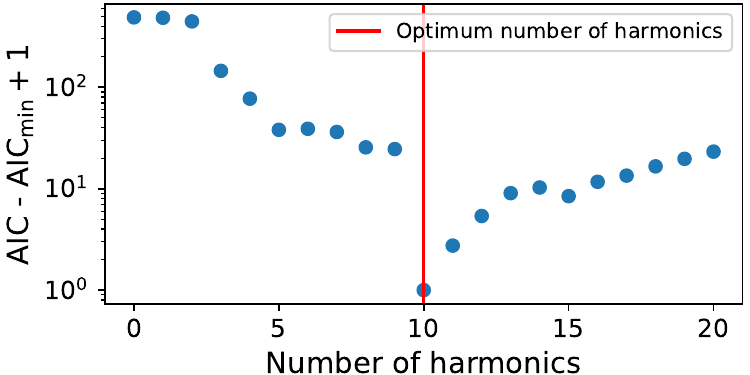}
&  \includegraphics[width=0.3\textwidth]{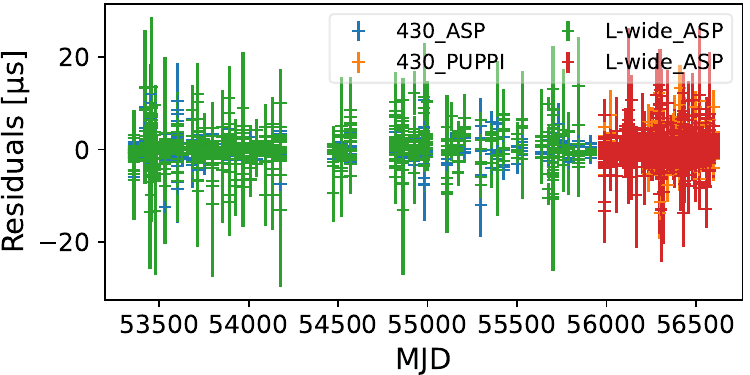}
     & \includegraphics[width=0.3\textwidth]{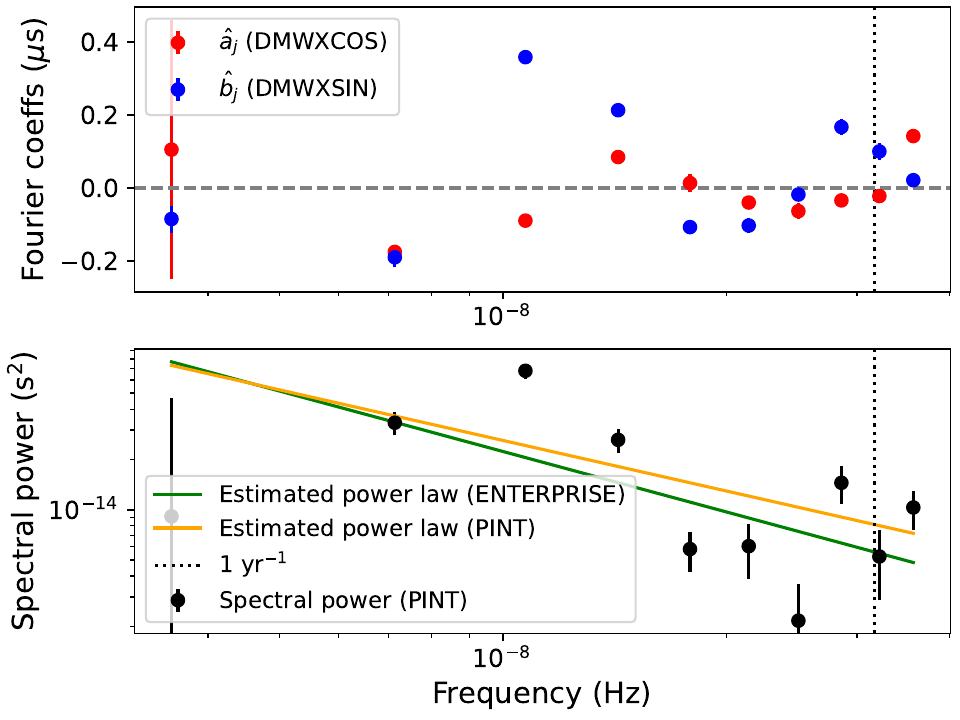} \\
    (a) & (b) & (c)
\end{tabular}       \\
     \includegraphics[scale=0.25]{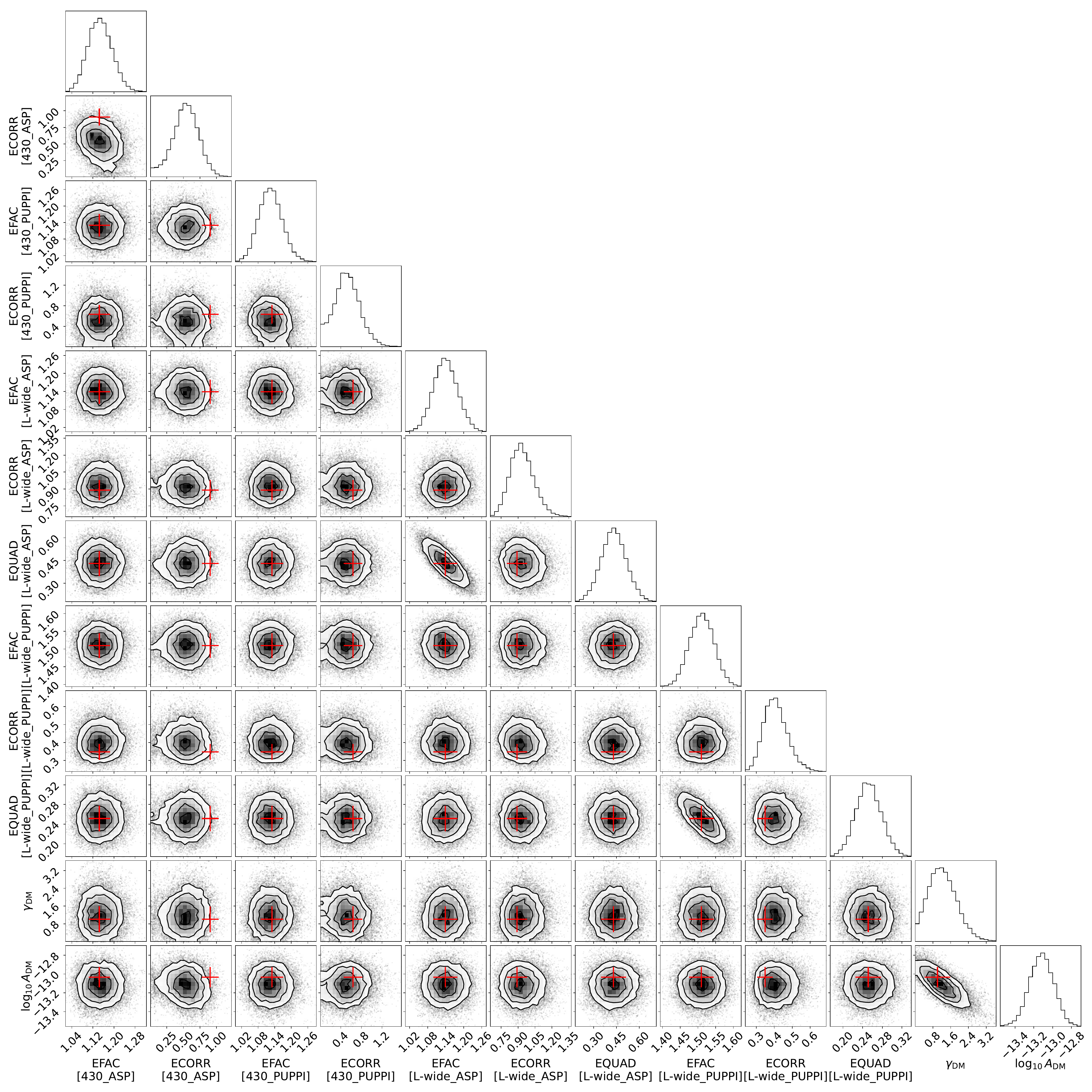}
     \\
     (d)
\end{tabular}  
}
\caption{Parameter estimation results for the NG9 dataset of PSR B1855+09 (see Section \ref{sec:b1855+09}).
\textit{Panel (a)} shows the $\text{AIC}-\text{AIC}_{\min} + 1$ values for the different number of DMN harmonics with identical plotting conventions to Figure \ref{fig:sim3}(a).  
\textit{Panel (b)} shows the post-fit timing residuals. The different colors in this panel represent different receiver-backend combinations. 
\textit{Panel (c)} shows the parameter estimation results for the DMN, with identical plotting conventions to Figure \ref{fig:sim3}(c).
\textit{Panel (d)} shows the comparison of the noise parameters obtained using \pint{} and \enterprise{}, with identical plotting conventions as Figure \ref{fig:sim1}.
Note that the EQUAD and ECORR parameters have units of $\mu$s.}
\label{fig:B1855}
\end{figure*}

We then used the AIC to determine which noise parameters are necessary for the given dataset as demonstrated in subsections \ref{sec:sim1} and \ref{sec:sim2}, and it turned out that EQUAD parameters were not needed for \textsl{430\_ASP} and \textsl{430\_PUPPI}. 
This was verified using the Savage-Dickey ratio \citep{Dickey1971} for these parameters, obtained from a Bayesian analysis performed using \enterprise{} and \ptmcmc.
Similarly, the optimal number of \texttt{DMWaveX} harmonics was determined to be 10 using the AIC similar to Subsection \ref{sec:sim4}, and this is shown in Figure \ref{fig:B1855}(a). 
\updated{The optimal number of \texttt{WaveX} harmonics turned out to be only 1, with an AIC difference of $\sim$16 against a model without \texttt{WaveX} but including \texttt{DMWaveX} and time-uncorrelated noise parameters as mentioned above.
The sine amplitude $a_i$ associated with this single harmonic turned out to be consistent with zero, and we found high covariances between these parameters and the rotational frequency and frequency derivative.
Hence, we excluded \texttt{WaveX} from our subsequent analysis.}

Finally, we performed a maximum-likelihood fit using the optimal model.
The spectral parameters of the DMN were estimated following Subsection \ref{sec:rednoisefit}.
The post-fit timing residuals are plotted in \ref{fig:B1855}(b), and the maximum-likelihood DMN parameter estimates are plotted in \ref{fig:B1855}(c).
Figure \ref{fig:B1855}(d) shows a comparison between the maximum-likelihood noise parameter estimates and their Bayesian counterparts (see Table \ref{tab:priors} for the prior distributions used) using a corner plot.
These plots show that the frequentist and Bayesian estimates agree with each other within their respective uncertainties.

\section{New features in \pint{}}
\label{sec:newdevs}

In this Section, we briefly summarize the new features that have been implemented in \pint{} since the publication of \citet{LuoRansom+2021}.
The new timing model components and new fitting methods were discussed in Sections \ref{sec:timing-model} and \ref{sec:timingfit}--\ref{sec:noisefit} respectively, and are not included here.

\subsection{Timing model comparison and conversion utilities}
\label{sec:convutils}

Different timing models can be compared with each other using the \texttt{TimingModel.compare()} function. 
This functionality is also available via the \texttt{compare\_parfiles} command line utility.

The \texttt{convert\_binary()} function in the \texttt{pint.binaryconvert} module allows the user to convert between different binary models (see Table \ref{tab:model-components}).
This functionality can also be accessed via the \texttt{convert\_parfile} command line utility.
\texttt{convert\_parfile} also allows the user to convert between different `par' file formats.

\pint{} only supports the TDB timescale internally unlike the \tempotwo{} package, which supports both TDB and TCB timescales \citep{HobbsEdwardsManchester2006}.
However, \pint{} can now read `par' files in the TCB timescale and convert them to TDB automatically.
This conversion can also be performed using the \texttt{tcb2tdb} command line utility.

\subsection{Global repository for clock files}
Since the TOAs are usually measured against local observatory clocks, a series of clock corrections must be applied to bring them \updated{to a particular realization of the TT timescale and then} to the TDB timescale (see \citet{LuoRansom+2021} for details).
These clock corrections are usually distributed as `clock files' containing \updated{a series of measured differences between the local observatory clock and} an international time standard (usually UTC(GPS)) over time.
\pint{} now accesses these files from a central repository\footnote{\url{https://github.com/ipta/pulsar-clock-corrections}} maintained by the International Pulsar Timing Array consortium \citep{VerbiestLentati+2016}.
This allows \pint{} to retain access to the most up-to-date clock corrections without the user having to manually update the clock files.

\subsection{TOA simulations}
\label{sec:pintsim}
\pint{}'s simulation functionality is implemented in the \texttt{pint.simulation} module, and can be invoked using the \texttt{zima} command line utility.
This module now provides a wide range of functionality on top of simple TOA generation, including the simulation of wideband TOAs, simulation of white noise incorporating EFACs and EQUADs, simulation of different types of single-pulsar correlated noise including ECORR, ARN, and DMN.
Note that this module can only simulate single-pulsar signals, and it cannot simulate signals that are correlated across pulsars such as the gravitational wave background.

\subsection{Bayesian interface}
The \texttt{BayesianTiming} class in the \texttt{pint.bayesian} module can be used to perform Bayesian parameter estimation and model selection for pulsar timing datasets.
This interface can be used to evaluate the likelihood function, prior distribution, and the prior transform function, and is compatible with both MCMC and nested samplers. 
This interface supports Bayesian inference on both narrowband and wideband datasets and allows the user to estimate the timing model and white noise parameters.
However, estimating correlated noise parameters (in their GP representation) is not yet supported.

\subsection{Chi-squared grids}
In some cases, rather than determine the best-fit values of all parameters, it is desirable to fit for all but a few parameters while stepping over a grid of a subset of the parameters.  
This is commonly done with post-Keplerian binary parameters \citep{DamourDeruelle1986} that can constrain the pulsar and companion masses \citep[e.g.,][]{CromartieFonseca+2020, FonsecaCromartie+2021}, since these parameters are often only marginally significant and give overlapping constraints on the masses.  

To facilitate this, we have implemented within the \texttt{pint.gridutils} module several methods. 
These enable the $\chi^2$  to be computed over grids of intrinsic parameters (such as Shapiro delay companion mass and range) or derived parameters created from combinations of intrinsic parameters (such as characteristic age).  
The individual model fits are done with the help of the \texttt{concurrent.futures} module \citep{Groot2020}  that enables asynchronously launching and tracking multiple jobs and has extensions like \texttt{clusterfutures}\footnote{\url{https://github.com/sampsyo/clusterfutures}} that enable deployment on high-performance computing clusters with batch scheduling using \texttt{slurm} \citep{YooJetteGrondona2003}.

\subsection{Publication output}
The \texttt{publish()} function in the \texttt{pint.output.publish} module can be used to generate a \latex{} table summary of a dataset comprising a timing model and a set of TOAs.
The same functionality is also available through the \texttt{pintpublish} command line utility.

\section{Summary and Discussion}
\label{sec:summary}

In this paper, we describe the new developments in the \pint{} pulsar timing package, with a focus on frequentist parameter estimation methods.
This includes the newly implemented timing model components (Table \ref{tab:model-components}), fitting algorithms (Table \ref{tab:fitters}), and features for performing specific tasks such as simulation, Bayesian inference, etc (Section \ref{sec:newdevs}).
In particular, we described the Downhill fitter algorithm, an improved version of the linear fitting algorithm commonly used in pulsar timing that is robust against significant non-linearity in the likelihood function as well as regions of the parameter space where the likelihood function is ill-defined.

We presented a new framework within the \pint{} pulsar timing package to perform frequentist pulsar timing noise characterization, involving the maximum-likelihood estimation of both uncorrelated and correlated noise parameters \updated{together} with timing model parameters, as well as a model comparison functionality using the AIC.
We demonstrated our parameter estimation and model comparison methods using simulated datasets as well as the NG9 dataset of PSR B1855+09.
The results obtained from these exercises show good agreement between our methods and conventional Bayesian methods, indicating the reliability of our methods. 
\updated{However, we find small parameter estimation biases (within $1\sigma$, but still statistically significant) in our simulations.
In future work, we plan to explore these biases and the various covariances that may arise due to data gaps, low observing bandwidth, etc.}

Additionally, we also present other new features of \pint{}, such as new timing model components, timing model comparison \& conversion utilities, a global repository for clock files, TOA simulation, an interface for Bayesian analysis, chi-squared grids, and a publication output utility.

The new noise characterization framework should improve the task of pulsar timing in two ways.
In pulsar timing projects where Bayesian analysis is deemed not worth the cost, noise characterization is either not done at all or is only done in an ad hoc manner.
In such projects, our new framework can provide a convenient and inexpensive alternative for noise characterization, ensuring more robust timing solutions.
In high-precision applications where Bayesian noise characterization is necessary, our framework can help accelerate the data preparation/combination stage by integrating noise characterization into interactive pulsar timing workflows and pulsar timing pipelines without needing a Bayesian analysis step.
The quick parameter estimates and model comparisons provided by our framework can quicken the iterative refinement of noise models.
Further, the frequentist noise parameter estimates can act as independent cross-checks for Bayesian results and as starting points for MCMC samplers to help them converge faster.

\section*{Acknowledgments}
This work has been carried out as part of the NANOGrav collaboration, which receives support from the National Science Foundation (NSF) Physics Frontiers Center award numbers 1430284 and 2020265.
Portions of this work performed at NRL were supported by ONR 6.1 basic research funding.
T.T.P. acknowledges support from the Extragalactic Astrophysics Research Group at E\"{o}tv\"{o}s Lor\'{a}nd University, funded by the E\"{o}tv\"{o}s Lor\'{a}nd Research Network (ELKH), which was used during the development of this research.
S.M.R. is a CIFAR Fellow.
M.B. was supported in part by  PRIN TEC INAF 2019 ``SpecTemPolar! -- Timing analysis in the era of high-throughput photon detectors’’, and CICLOPS -- Citizen Computing Pulsar Search, a project supported by \textit{POR FESR Sardegna 2014 – 2020 Asse 1 Azione 1.1.3} (code RICERCA\_1C-181), call for proposal "Aiuti per Progetti di Ricerca e Sviluppo 2017" managed by Sardegna Ricerche.
H.T.C. acknowledges funding from the U.S. Naval Research Laboratory.
A.S. thanks David Nice and Bhal Chandra Joshi for helpful discussions.

%

\vspace{5mm}


\software{
\texttt{numpy} \citep{HarrisMillman+2020}, 
\texttt{scipy} \citep{VirtanenGommers+2020}, 
\texttt{astropy} \citep{Price-WhelanLim+2022}, 
\texttt{matplotlib} \citep{Hunter2007}, 
\texttt{corner} \citep{Foreman-Mackey2016}, 
\texttt{numdifftools}\footnote{\url{https://github.com/pbrod/numdifftools}}, 
\pint{} \citep{LuoRansom+2021}, 
\enterprise{} \citep{JohnsonMeyers+2024}, 
\texttt{enterprise\_extensions} \citep{JohnsonMeyers+2024}, 
\texttt{PTMCMCSampler} \citep{JohnsonMeyers+2024}, 
\texttt{emcee} \citep{Foreman-MackeyHogg+2013}, 
\texttt{Jupyter}\footnote{\url{https://jupyter.org/}}.
}

~

\textit{\large Data Availability:}
The NANOGrav 9-year dataset of PSR B1855+09 is available at \url{https://nanograv.org/science/data/9-year-pulsar-timing-array-data-release}.
It is also distributed as an example dataset along with the \pint{} source code at \url{https://github.com/nanograv/PINT}.
The simulated datasets used in this paper and the Jupyter notebooks used to create and analyze them are available at \url{https://github.com/abhisrkckl/pint-noise}.



\appendix

\section{System and frequency-dependent delays}
\label{sec:fdjump}

The phenomenological model used to account for frequency-profile evolution is given by
\begin{equation}
    \Delta_\text{FD} = \sum_p \mathfrak{F}_p \left[\log\left(\frac{\nu}{\text{1 GHz}}\right)\right]^p\,,
\end{equation}
where $p$ is an index and $\mathfrak{F}_p$ is known as an FD parameter.
In large data sets containing TOAs obtained using many different observing systems, it is often insufficient to model the frequency-dependent profile evolution using global FD parameters that affect all TOAs.
The need for system-dependent FD parameters can arise due to (a) the data reduction procedures used for different observing systems being different, (b) different template profiles being used to compute TOAs for different systems, etc.
The system-dependent FD delay is given by
\begin{equation}
    \Delta_\text{FDJump} = \sum_{pq} \mathfrak{F}_{pq} \left[\log\left(\frac{\nu}{\text{1 GHz}}\right)\right]^p \mathcal{M}_{q}\,,
    \label{eq:fdjump_log}
\end{equation}
where $\mathfrak{F}_{pq}$ is known as a system-dependent FD parameter (also known as an FD jump), and $\mathcal{M}_{q}$ is a TOA selection mask.
An alternative model for $\Delta_\text{FDJump}$, originally implemented in \tempotwo{}, is also available:
\begin{equation}
    \Delta_\text{FDJump} = \sum_{pq} \mathfrak{F}_{pq} \left(\frac{\nu}{\text{1 GHz}}\right)^p \mathcal{M}_{q}\,.
    \label{eq:fdjump_nolog}
\end{equation}
The two expressions above can be toggled using a Boolean parameter \texttt{FDJUMPLOG}.

\bibliographystyle{aasjournal}
\bibliography{pint-noise}{}

\begin{thebibliography}{}
\expandafter\ifx\csname natexlab\endcsname\relax\def\natexlab#1{#1}\fi
\providecommand{\url}[1]{\href{#1}{#1}}
\providecommand{\dodoi}[1]{doi:~\href{http://doi.org/#1}{\nolinkurl{#1}}}
\providecommand{\doeprint}[1]{\href{http://ascl.net/#1}{\nolinkurl{http://ascl.net/#1}}}
\providecommand{\doarXiv}[1]{\href{https://arxiv.org/abs/#1}{\nolinkurl{https://arxiv.org/abs/#1}}}

\bibitem[{Agazie {et~al.}(2024)Agazie, Antoniadis, Anumarlapudi, Archibald, Arumugam, {et~al.}}]{AgazieAntoniadis+2024}
Agazie, G., Antoniadis, J., Anumarlapudi, A., {et~al.} 2024, The Astrophysical Journal, 966, 105, \dodoi{10.3847/1538-4357/ad36be}

\bibitem[{Agazie {et~al.}(2023{\natexlab{a}})Agazie, Anumarlapudi, Archibald, Arzoumanian, Baker, {et~al.}}]{AgazieAnumarlapudi+2023_ng15detchar}
Agazie, G., Anumarlapudi, A., Archibald, A.~M., {et~al.} 2023{\natexlab{a}}, The Astrophysical Journal Letters, 951, L10, \dodoi{10.3847/2041-8213/acda88}

\bibitem[{Agazie {et~al.}(2023{\natexlab{b}})Agazie, Anumarlapudi, Archibald, Baker, Bécsy, {et~al.}}]{AgazieAnumarlapudi+2023_ng15gwb}
---. 2023{\natexlab{b}}, The Astrophysical Journal Letters, 952, L37, \dodoi{10.3847/2041-8213/ace18b}

\bibitem[{{Alam} {et~al.}(2021){Alam}, {Arzoumanian}, {Baker}, {Blumer}, {Bohler}, {et~al.}}]{AlamArzoumanian+2021}
{Alam}, M.~F., {Arzoumanian}, Z., {Baker}, P.~T., {et~al.} 2021, The Astrophysical Journal Supplement Series, 252, 5, \dodoi{10.3847/1538-4365/abc6a1}

\bibitem[{{Antoniadis} {et~al.}(2023){Antoniadis}, {Arumugam}, {Arumugam}, {Babak}, {Bagchi}, {et~al.}}]{AntoniadisArumugam+2023_eptadr2gwb}
{Antoniadis}, J., {Arumugam}, P., {Arumugam}, S., {et~al.} 2023, Astronomy \& Astrophysics, 678, A50, \dodoi{10.1051/0004-6361/202346844}

\bibitem[{Arzoumanian {et~al.}(2015)Arzoumanian, Brazier, Burke-Spolaor, Chamberlin, Chatterjee, {et~al.}}]{ArzoumanianBrazier+2015}
Arzoumanian, Z., Brazier, A., Burke-Spolaor, S., {et~al.} 2015, The Astrophysical Journal, 813, 65, \dodoi{10.1088/0004-637X/813/1/65}

\bibitem[{{Ashton} {et~al.}(2022){Ashton}, {Bernstein}, {Buchner}, {Chen}, {Cs{\'a}nyi}, {Fowlie}, {Feroz}, {Griffiths}, {Handley}, {Habeck}, {Higson}, {Hobson}, {Lasenby}, {Parkinson}, {P{\'a}rtay}, {Pitkin}, {Schneider}, {Speagle}, {South}, {Veitch}, {Wacker}, {Wales}, \& {Yallup}}]{AshtonBernstein+2022}
{Ashton}, G., {Bernstein}, N., {Buchner}, J., {et~al.} 2022, Nature Reviews Methods Primers, 2, 39, \dodoi{10.1038/s43586-022-00121-x}

\bibitem[{{Backer} \& {Hellings}(1986)}]{BackerHellings1986}
{Backer}, D.~C., \& {Hellings}, R.~W. 1986, Annual Review of Astronomy and Astrophysics, 24, 537, \dodoi{10.1146/annurev.aa.24.090186.002541}

\bibitem[{{Blandford} {et~al.}(1984){Blandford}, {Narayan}, \& {Romani}}]{BlandfordNarayanRomani1984}
{Blandford}, R., {Narayan}, R., \& {Romani}, R.~W. 1984, Journal of Astrophysics and Astronomy, 5, 369, \dodoi{10.1007/BF02714466}

\bibitem[{{Blandford} \& {Teukolsky}(1976)}]{BlandfordTeukolsky1976}
{Blandford}, R., \& {Teukolsky}, S.~A. 1976, The Astrophysical Journal, 205, 580, \dodoi{10.1086/154315}

\bibitem[{Burnham \& Anderson(2004)}]{BurnhamAnderson2004}
Burnham, K.~P., \& Anderson, D.~R. 2004, Sociological Methods \& Research, 33, 261, \dodoi{10.1177/0049124104268644}

\bibitem[{Caballero {et~al.}(2018)Caballero, Guo, Lee, Lazarus, Champion, {et~al.}}]{CaballeroGuo+2018}
Caballero, R.~N., Guo, Y.~J., Lee, K.~J., {et~al.} 2018, Monthly Notices of the Royal Astronomical Society, 481, 5501, \dodoi{10.1093/mnras/sty2632}

\bibitem[{Coles {et~al.}(2011)Coles, Hobbs, Champion, Manchester, \& Verbiest}]{ColesHobbs+2011}
Coles, W., Hobbs, G., Champion, D.~J., Manchester, R.~N., \& Verbiest, J. P.~W. 2011, Monthly Notices of the Royal Astronomical Society, 418, 561, \dodoi{10.1111/j.1365-2966.2011.19505.x}

\bibitem[{{Cromartie} {et~al.}(2020){Cromartie}, {Fonseca}, {Ransom}, {Demorest}, {Arzoumanian}, {et~al.}}]{CromartieFonseca+2020}
{Cromartie}, H.~T., {Fonseca}, E., {Ransom}, S.~M., {et~al.} 2020, Nature Astronomy, 4, 72, \dodoi{10.1038/s41550-019-0880-2}

\bibitem[{{Damour} \& {Deruelle}(1986)}]{DamourDeruelle1986}
{Damour}, T., \& {Deruelle}, N. 1986, Annales de L'Institut Henri Poincare Section (A) Physique Theorique, 44, 263.
\newblock \url{http://www.numdam.org/item/AIHPA_1986__44_3_263_0/}

\bibitem[{Damour \& Taylor(1992)}]{DamourTaylor1992}
Damour, T., \& Taylor, J.~H. 1992, Physical Review D, 45, 1840, \dodoi{10.1103/PhysRevD.45.1840}

\bibitem[{{Davis} {et~al.}(1985){Davis}, {Herring}, {Shapiro}, {Rogers}, \& {Elgered}}]{DavisHerring+1985}
{Davis}, J.~L., {Herring}, T.~A., {Shapiro}, I.~I., {Rogers}, A.~E.~E., \& {Elgered}, G. 1985, Radio Science, 20, 1593, \dodoi{10.1029/RS020i006p01593}

\bibitem[{{de Groot}(2020)}]{Groot2020}
{de Groot}, C. 2020, The Concurrent.Futures Library (Apress), 12, \dodoi{10.1007/978-1-4842-6582-6_12}

\bibitem[{Demorest {et~al.}(2012)Demorest, Ferdman, Gonzalez, Nice, Ransom, {et~al.}}]{DemorestFerdman+2013}
Demorest, P.~B., Ferdman, R.~D., Gonzalez, M.~E., {et~al.} 2012, The Astrophysical Journal, 762, 94, \dodoi{10.1088/0004-637X/762/2/94}

\bibitem[{Deng {et~al.}(2012)Deng, Coles, Hobbs, Keith, Manchester, Shannon, \& Zheng}]{DengColes+2012}
Deng, X.~P., Coles, W.~A., Hobbs, G.~B., {et~al.} 2012, Monthly Notices of the Royal Astronomical Society, 424, 244, \dodoi{10.1111/j.1365-2966.2012.21189.x}

\bibitem[{Dickey(1971)}]{Dickey1971}
Dickey, J.~M. 1971, The Annals of Mathematical Statistics, 42, 204 , \dodoi{10.1214/aoms/1177693507}

\bibitem[{{Donner} {et~al.}(2020){Donner}, {Verbiest}, {Tiburzi}, {Os{\l}owski}, {K{\"u}nsem{\"o}ller}, {et~al.}}]{DonnerVerbiest+2020}
{Donner}, J.~Y., {Verbiest}, J.~P.~W., {Tiburzi}, C., {et~al.} 2020, Astronomy \& Astrophysics, 644, A153, \dodoi{10.1051/0004-6361/202039517}

\bibitem[{Edwards {et~al.}(2006)Edwards, Hobbs, \& Manchester}]{EdwardsHobbsManchester2006}
Edwards, R.~T., Hobbs, G.~B., \& Manchester, R.~N. 2006, Monthly Notices of the Royal Astronomical Society, 372, 1549, \dodoi{10.1111/j.1365-2966.2006.10870.x}

\bibitem[{Fiore {et~al.}(2023)Fiore, Levin, McLaughlin, Anumarlapudi, Kaplan, {et~al.}}]{FioreLevin+2023}
Fiore, W., Levin, L., McLaughlin, M.~A., {et~al.} 2023, The Astrophysical Journal, 956, 40, \dodoi{10.3847/1538-4357/aceef7}

\bibitem[{{Fonseca} {et~al.}(2021){Fonseca}, {Cromartie}, {Pennucci}, {Ray}, {Kirichenko}, {Ransom}, {Demorest}, {Stairs}, {Arzoumanian}, {Guillemot}, {Parthasarathy}, {Kerr}, {Cognard}, {Baker}, {Blumer}, {Brook}, {DeCesar}, {Dolch}, {Dong}, {Ferrara}, {Fiore}, {Garver-Daniels}, {Good}, {Jennings}, {Jones}, {Kaspi}, {Lam}, {Lorimer}, {Luo}, {McEwen}, {McKee}, {McLaughlin}, {McMann}, {Meyers}, {Naidu}, {Ng}, {Nice}, {Pol}, {Radovan}, {Shapiro-Albert}, {Tan}, {Tendulkar}, {Swiggum}, {Wahl}, \& {Zhu}}]{FonsecaCromartie+2021}
{Fonseca}, E., {Cromartie}, H.~T., {Pennucci}, T.~T., {et~al.} 2021, The Astrophysical Journal Letters, 915, L12, \dodoi{10.3847/2041-8213/ac03b8}

\bibitem[{Foreman-Mackey(2016)}]{Foreman-Mackey2016}
Foreman-Mackey, D. 2016, The Journal of Open Source Software, 1, 24, \dodoi{10.21105/joss.00024}

\bibitem[{Foreman-Mackey {et~al.}(2013)Foreman-Mackey, Hogg, Lang, \& Goodman}]{Foreman-MackeyHogg+2013}
Foreman-Mackey, D., Hogg, D.~W., Lang, D., \& Goodman, J. 2013, Publications of the Astronomical Society of the Pacific, 125, 306, \dodoi{10.1086/670067}

\bibitem[{Foster \& Backer(1990)}]{FosterBacker1990}
Foster, R.~S., \& Backer, D.~C. 1990, The Astrophysical Journal, 361, 300, \dodoi{10.1086/169195}

\bibitem[{{Freire} \& {Wex}(2010)}]{FreireWex2010}
{Freire}, P. C.~C., \& {Wex}, N. 2010, Monthly Notices of the Royal Astronomical Society, 409, 199, \dodoi{10.1111/j.1365-2966.2010.17319.x}

\bibitem[{{Harris} {et~al.}(2020){Harris}, {Millman}, {van der Walt}, {Gommers}, {Virtanen}, {et~al.}}]{HarrisMillman+2020}
{Harris}, C.~R., {Millman}, K.~J., {van der Walt}, S.~J., {et~al.} 2020, Nature, 585, 357, \dodoi{10.1038/s41586-020-2649-2}

\bibitem[{Hazboun {et~al.}(2022)Hazboun, Simon, Madison, Arzoumanian, Cromartie, {et~al.}}]{HazbounSimon+2022}
Hazboun, J.~S., Simon, J., Madison, D.~R., {et~al.} 2022, The Astrophysical Journal, 929, 39, \dodoi{10.3847/1538-4357/ac5829}

\bibitem[{Hee {et~al.}(2015)Hee, Handley, Hobson, \& Lasenby}]{HeeHandley+2015}
Hee, S., Handley, W.~J., Hobson, M.~P., \& Lasenby, A.~N. 2015, Monthly Notices of the Royal Astronomical Society, 455, 2461, \dodoi{10.1093/mnras/stv2217}

\bibitem[{Hobbs(2014)}]{Hobbs2014}
Hobbs, G. 2014, TEMPO2 examples.
\newblock \url{https://www.jb.man.ac.uk/~pulsar/Resources/tempo2_examples_ver1.pdf}

\bibitem[{Hobbs {et~al.}(2019)Hobbs, Guo, Caballero, Coles, Lee, {et~al.}}]{HobbsGuo+2019}
Hobbs, G., Guo, L., Caballero, R.~N., {et~al.} 2019, Monthly Notices of the Royal Astronomical Society, 491, 5951, \dodoi{10.1093/mnras/stz3071}

\bibitem[{{Hobbs} {et~al.}(2006){Hobbs}, {Edwards}, \& {Manchester}}]{HobbsEdwardsManchester2006}
{Hobbs}, G.~B., {Edwards}, R.~T., \& {Manchester}, R.~N. 2006, Monthly Notices of the Royal Astronomical Society, 369, 655, \dodoi{10.1111/j.1365-2966.2006.10302.x}

\bibitem[{{Hotan} {et~al.}(2004){Hotan}, {van Straten}, \& {Manchester}}]{HotanVanStratenManchester2004}
{Hotan}, A.~W., {van Straten}, W., \& {Manchester}, R.~N. 2004, Publications of the Astronomical Society of Australia, 21, 302, \dodoi{10.1071/AS04022}

\bibitem[{Hunter(2007)}]{Hunter2007}
Hunter, J.~D. 2007, Computing in Science \& Engineering, 9, 90, \dodoi{10.1109/MCSE.2007.55}

\bibitem[{{Johnson} {et~al.}(2024){Johnson}, {Meyers}, {Baker}, {Cornish}, {Hazboun}, {et~al.}}]{JohnsonMeyers+2024}
{Johnson}, A.~D., {Meyers}, P.~M., {Baker}, P.~T., {et~al.} 2024, Physical Review D, 109, 103012, \dodoi{10.1103/PhysRevD.109.103012}

\bibitem[{Jones \& Qin(2022)}]{JonesQin2022}
Jones, G.~L., \& Qin, Q. 2022, Annual Review of Statistics and Its Application, 9, 557, \dodoi{10.1146/annurev-statistics-040220-090158}

\bibitem[{Keith \& Niţu(2023)}]{KeithNitu2023}
Keith, M.~J., \& Niţu, I.~C. 2023, Monthly Notices of the Royal Astronomical Society, 523, 4603, \dodoi{10.1093/mnras/stad1713}

\bibitem[{{Kopeikin}(1995)}]{Kopeikin1995}
{Kopeikin}, S.~M. 1995, The Astrophysical Journal Letters, 439, L5, \dodoi{10.1086/187731}

\bibitem[{{Kopeikin}(1996)}]{Kopeikin1996}
---. 1996, The Astrophysical Journal Letters, 467, L93, \dodoi{10.1086/310201}

\bibitem[{{Kramer} {et~al.}(2006){Kramer}, {Stairs}, {Manchester}, {McLaughlin}, {Lyne}, {et~al.}}]{KramerStairs+2006}
{Kramer}, M., {Stairs}, I.~H., {Manchester}, R.~N., {et~al.} 2006, Science, 314, 97, \dodoi{10.1126/science.1132305}

\bibitem[{Kramer {et~al.}(2021)Kramer, Stairs, Manchester, Wex, Deller, {et~al.}}]{KramerStairs+2021}
Kramer, M., Stairs, I.~H., Manchester, R.~N., {et~al.} 2021, Physical Review X, 11, 041050, \dodoi{10.1103/PhysRevX.11.041050}

\bibitem[{Laal {et~al.}(2023)Laal, Lamb, Romano, Siemens, Taylor, \& van Haasteren}]{LaalLamb+2023}
Laal, N., Lamb, W.~G., Romano, J.~D., {et~al.} 2023, Physical Review D, 108, 063008, \dodoi{10.1103/PhysRevD.108.063008}

\bibitem[{Lange {et~al.}(2001)Lange, Camilo, Wex, Kramer, Backer, Lyne, \& Doroshenko}]{LangeCamilo+2001}
Lange, C., Camilo, F., Wex, N., {et~al.} 2001, Monthly Notices of the Royal Astronomical Society, 326, 274, \dodoi{10.1046/j.1365-8711.2001.04606.x}

\bibitem[{{Lentati} {et~al.}(2014){Lentati}, {Alexander}, {Hobson}, {Feroz}, {van Haasteren}, {Lee}, \& {Shannon}}]{LentatiHobson+2014}
{Lentati}, L., {Alexander}, P., {Hobson}, M.~P., {et~al.} 2014, Monthly Notices of the Royal Astronomical Society, 437, 3004, \dodoi{10.1093/mnras/stt2122}

\bibitem[{{Lorimer} \& {Kramer}(2012)}]{LorimerKramer2012}
{Lorimer}, D.~R., \& {Kramer}, M. 2012, {Handbook of Pulsar Astronomy} (Cambridge University Press)

\bibitem[{{Luo} {et~al.}(2021){Luo}, {Ransom}, {Demorest}, {Ray}, {Archibald}, {et~al.}}]{LuoRansom+2021}
{Luo}, J., {Ransom}, S., {Demorest}, P., {et~al.} 2021, The Astrophysical Journal, 911, 45, \dodoi{10.3847/1538-4357/abe62f}

\bibitem[{{Madison} {et~al.}(2019){Madison}, {Cordes}, {Arzoumanian}, {Chatterjee}, {Crowter}, {et~al.}}]{MadisonCordes+2019}
{Madison}, D.~R., {Cordes}, J.~M., {Arzoumanian}, Z., {et~al.} 2019, The Astrophysical Journal, 872, 150, \dodoi{10.3847/1538-4357/ab01fd}

\bibitem[{{Manchester}(2017)}]{Manchester2017}
{Manchester}, R.~N. 2017, Journal of Astrophysics and Astronomy, 38, 42, \dodoi{10.1007/s12036-017-9469-2}

\bibitem[{{Nice} {et~al.}(2015){Nice}, {Demorest}, {Stairs}, {Manchester}, {Taylor}, {Peters}, {et~al.}}]{NiceDemorest+2015}
{Nice}, D., {Demorest}, P., {Stairs}, I., {et~al.} 2015, {Tempo: Pulsar timing data analysis}, Astrophysics Source Code Library, record ascl:1509.002.
\newblock \doeprint{1509.002}

\bibitem[{{Niell}(1996)}]{Niell1996}
{Niell}, A.~E. 1996, Journal of Geophysical Research, 101, 3227, \dodoi{10.1029/95JB03048}

\bibitem[{{Park} {et~al.}(2021){Park}, {Folkner}, {Williams}, \& {Boggs}}]{ParkFolkner+2021}
{Park}, R.~S., {Folkner}, W.~M., {Williams}, J.~G., \& {Boggs}, D.~H. 2021, The Astronomical Journal, 161, 105, \dodoi{10.3847/1538-3881/abd414}

\bibitem[{Pennucci(2019)}]{Pennucci2019}
Pennucci, T.~T. 2019, The Astrophysical Journal, 871, 34, \dodoi{10.3847/1538-4357/aaf6ef}

\bibitem[{{Pennucci} {et~al.}(2014){Pennucci}, {Demorest}, \& {Ransom}}]{PennucciDemorestRansom2014}
{Pennucci}, T.~T., {Demorest}, P.~B., \& {Ransom}, S.~M. 2014, The Astrophysical Journal, 790, 93, \dodoi{10.1088/0004-637X/790/2/93}

\bibitem[{Powell(1964)}]{Powell1964}
Powell, M. J.~D. 1964, The Computer Journal, 7, 155, \dodoi{10.1093/comjnl/7.2.155}

\bibitem[{Press {et~al.}(1992)Press, Teukolsky, Vetterling, \& Flannery}]{PressTeukolsky+1992}
Press, W.~H., Teukolsky, S.~A., Vetterling, W.~T., \& Flannery, B.~P. 1992, Numerical Recipes in C: The Art of Scientific Computing, 2nd edn. (Cambridge University Press), 32--102

\bibitem[{Price-Whelan {et~al.}(2022)Price-Whelan, Lim, Earl, Starkman, Bradley, {et~al.}}]{Price-WhelanLim+2022}
Price-Whelan, A.~M., Lim, P.~L., Earl, N., {et~al.} 2022, The Astrophysical Journal, 935, 167, \dodoi{10.3847/1538-4357/ac7c74}

\bibitem[{Rafikov \& Lai(2006)}]{RafikovLai2006}
Rafikov, R.~R., \& Lai, D. 2006, Physical Review D, 73, 063003, \dodoi{10.1103/PhysRevD.73.063003}

\bibitem[{{Ransom}(2001)}]{Ransom2001}
{Ransom}, S.~M. 2001, PhD thesis, Harvard University, Massachusetts

\bibitem[{{Ray} {et~al.}(2019){Ray}, {Guillot}, {Ho}, {Kerr}, {Enoto}, {Gendreau}, {Arzoumanian}, {Altamirano}, {Bogdanov}, {Campion}, {Chakrabarty}, {Deneva}, {Jaisawal}, {Kozon}, {Malacaria}, {Strohmayer}, \& {Wolff}}]{RayGuillot+2019}
{Ray}, P.~S., {Guillot}, S., {Ho}, W. C.~G., {et~al.} 2019, The Astrophysical Journal, 879, 130, \dodoi{10.3847/1538-4357/ab24d8}

\bibitem[{Reardon {et~al.}(2023)Reardon, Zic, Shannon, Hobbs, Bailes, {et~al.}}]{ReardonZic+2023}
Reardon, D.~J., Zic, A., Shannon, R.~M., {et~al.} 2023, The Astrophysical Journal Letters, 951, L6, \dodoi{10.3847/2041-8213/acdd02}

\bibitem[{{Sazhin}(1978)}]{Sazhin1978}
{Sazhin}, M.~V. 1978, Soviet Astronomy, 22, 36

\bibitem[{Shapiro(1964)}]{Shapiro1964}
Shapiro, I.~I. 1964, Phys. Rev. Lett., 13, 789, \dodoi{10.1103/PhysRevLett.13.789}

\bibitem[{{Susobhanan} {et~al.}(2018){Susobhanan}, {Gopakumar}, {Joshi}, \& {Kumar}}]{SusobhananGopakumar+2018}
{Susobhanan}, A., {Gopakumar}, A., {Joshi}, B.~C., \& {Kumar}, R. 2018, Monthly Notices of the Royal Astronomical Society, 480, 5260, \dodoi{10.1093/mnras/sty2177}

\bibitem[{Tarafdar {et~al.}(2022)Tarafdar, Nobleson, Rana, Singha, Krishnakumar, {et~al.}}]{TarafdarNobleson+2022}
Tarafdar, P., Nobleson, K., Rana, P., {et~al.} 2022, Publications of the Astronomical Society of Australia, 39, e053, \dodoi{10.1017/pasa.2022.46}

\bibitem[{Taylor(1992)}]{Taylor1992}
Taylor, J.~H. 1992, Philosophical Transactions of the Royal Society of London. Series A: Physical and Engineering Sciences, 341, 117, \dodoi{10.1098/rsta.1992.0088}

\bibitem[{{Taylor} \& {Weisberg}(1989)}]{TaylorWeisberg1989}
{Taylor}, J.~H., \& {Weisberg}, J.~M. 1989, The Astrophysical Journal, 345, 434, \dodoi{10.1086/167917}

\bibitem[{{Tiburzi} {et~al.}(2021){Tiburzi}, {Shaifullah}, {Bassa}, {Zucca}, {Verbiest}, {et~al.}}]{TiburziShaifullah+2021}
{Tiburzi}, C., {Shaifullah}, G.~M., {Bassa}, C.~G., {et~al.} 2021, Astronomy \& Astrophysics, 647, A84, \dodoi{10.1051/0004-6361/202039846}

\bibitem[{{Vallisneri}(2020)}]{Vallisneri2020}
{Vallisneri}, M. 2020, {libstempo: Python wrapper for Tempo2}, Astrophysics Source Code Library, record ascl:2002.017.
\newblock \doeprint{2002.017}

\bibitem[{Vallisneri {et~al.}(2020)Vallisneri, Taylor, Simon, Folkner, Park, Cutler, Ellis, Lazio, Vigeland, Aggarwal, Arzoumanian, Baker, Brazier, Brook, Burke-Spolaor, Chatterjee, Cordes, Cornish, Crawford, Cromartie, Crowter, DeCesar, Demorest, Dolch, Ferdman, Ferrara, Fonseca, Garver-Daniels, Gentile, Good, Hazboun, Holgado, Huerta, Islo, Jennings, Jones, Jones, Kaplan, Kelley, Key, Lam, Levin, Lorimer, Luo, Lynch, Madison, McLaughlin, McWilliams, Mingarelli, Ng, Nice, Pennucci, Pol, Ransom, Ray, Siemens, Spiewak, Stairs, Stinebring, Stovall, Swiggum, van Haasteren, Witt, \& Zhu}]{VallisneriTaylor+2020}
Vallisneri, M., Taylor, S.~R., Simon, J., {et~al.} 2020, The Astrophysical Journal, 893, 112, \dodoi{10.3847/1538-4357/ab7b67}

\bibitem[{van Haasteren \& Levin(2012)}]{vanHaasterenLevin2012}
van Haasteren, R., \& Levin, Y. 2012, Monthly Notices of the Royal Astronomical Society, 428, 1147, \dodoi{10.1093/mnras/sts097}

\bibitem[{van Haasteren \& Vallisneri(2014)}]{VanHaasterenVallisneri2014a}
van Haasteren, R., \& Vallisneri, M. 2014, Monthly Notices of the Royal Astronomical Society, 446, 1170, \dodoi{10.1093/mnras/stu2157}

\bibitem[{{van Haasteren} \& {Vallisneri}(2014)}]{VanHaasterenVallisneri2014b}
{van Haasteren}, R., \& {Vallisneri}, M. 2014, Physical Review D, 90, 104012, \dodoi{10.1103/PhysRevD.90.104012}

\bibitem[{van Straten \& Bailes(2011)}]{vanStratenBailes2011}
van Straten, W., \& Bailes, M. 2011, Publications of the Astronomical Society of Australia, 28, 1–14, \dodoi{10.1071/AS10021}

\bibitem[{Verbiest {et~al.}(2016)Verbiest, Lentati, Hobbs, van Haasteren, Demorest, {et~al.}}]{VerbiestLentati+2016}
Verbiest, J. P.~W., Lentati, L., Hobbs, G., {et~al.} 2016, Monthly Notices of the Royal Astronomical Society, 458, 1267, \dodoi{10.1093/mnras/stw347}

\bibitem[{{Virtanen} {et~al.}(2020){Virtanen}, {Gommers}, {Oliphant}, {Haberland}, {Reddy}, {et~al.}}]{VirtanenGommers+2020}
{Virtanen}, P., {Gommers}, R., {Oliphant}, T.~E., {et~al.} 2020, Nature Methods, 17, 261, \dodoi{10.1038/s41592-019-0686-2}

\bibitem[{{Weisberg} \& {Huang}(2016)}]{WeisbergHuang2016}
{Weisberg}, J.~M., \& {Huang}, Y. 2016, The Astrophysical Journal, 829, 55, \dodoi{10.3847/0004-637X/829/1/55}

\bibitem[{{Wolszczan} \& {Frail}(1992)}]{WolszczanFrail1992}
{Wolszczan}, A., \& {Frail}, D.~A. 1992, Nature, 355, 145, \dodoi{10.1038/355145a0}

\bibitem[{Xu {et~al.}(2023)Xu, Chen, Guo, Jiang, Wang, {et~al.}}]{XuChen+2023}
Xu, H., Chen, S., Guo, Y., {et~al.} 2023, Research in Astronomy and Astrophysics, 23, 075024, \dodoi{10.1088/1674-4527/acdfa5}

\bibitem[{Yoo {et~al.}(2003)Yoo, Jette, \& Grondona}]{YooJetteGrondona2003}
Yoo, A.~B., Jette, M.~A., \& Grondona, M. 2003, in Job Scheduling Strategies for Parallel Processing, ed. D.~Feitelson, L.~Rudolph, \& U.~Schwiegelshohn (Berlin, Heidelberg: Springer Berlin Heidelberg), 44--60

\bibitem[{You {et~al.}(2012)You, Coles, Hobbs, \& Manchester}]{YouColes+2012}
You, X.~P., Coles, W.~A., Hobbs, G.~B., \& Manchester, R.~N. 2012, Monthly Notices of the Royal Astronomical Society, 422, 1160, \dodoi{10.1111/j.1365-2966.2012.20688.x}

\bibitem[{You {et~al.}(2007)You, Hobbs, Coles, Manchester, \& Han}]{YouHobbs+2007}
You, X.~P., Hobbs, G.~B., Coles, W.~A., Manchester, R.~N., \& Han, J.~L. 2007, The Astrophysical Journal, 671, 907, \dodoi{10.1086/522227}

\bibitem[{Zhu {et~al.}(2018)Zhu, Desvignes, Wex, Caballero, Champion, {et~al.}}]{ZhuDesvignes+2018}
Zhu, W.~W., Desvignes, G., Wex, N., {et~al.} 2018, Monthly Notices of the Royal Astronomical Society, 482, 3249, \dodoi{10.1093/mnras/sty2905}

\end{thebibliography}



\end{document}